\documentclass[%
twocolumn,
reprint,
 amsmath,amssymb,
 aps, physrev,
 prd,
]{revtex4-2}

\usepackage{graphicx}
\usepackage{dcolumn}

\usepackage{float}
\usepackage{subfigure}

\usepackage{braket}
\usepackage{amsmath}
\usepackage{amssymb}
\usepackage{array}
\usepackage{mathtools}
\usepackage{bm}
\usepackage{bbm}

\usepackage{tikz}
\usetikzlibrary{quantikz2}
\usetikzlibrary{3d} 

\usepackage{hyperref}

\usepackage{lipsum}

\newcommand{\DD}[2][]{\dfrac{\partial #1}{\partial #2}}

\renewcommand{\mod}[1]{\ \left(\mathrm{mod} \ #1\right)}

\newcommand{\I}{\mathbbm{\hat 1}}
\newcommand{\X}{\hat X}
\newcommand{\Z}{\hat Z}

\newcommand{\norm}[1]{\left\lVert#1\right\rVert}

\begin{document}


\title{A novel quantum lattice Boltzmann methodology for cosmological radiation transport}

\author{Rasmit Devkota}
\email{rdevkota3@gatech.edu}
\affiliation{
 School of Physics, Georgia Institute of Technology, Atlanta, GA 30332, USA
}
\affiliation{
 School of Mathematics, Georgia Institute of Technology, Atlanta, GA 30332, USA
}
 
\author{John H. Wise}
\email{jwise@physics.gatech.edu}
\affiliation{
 Center for Relativistic Astrophysics, School of Physics, Georgia Institute of Technology, Atlanta, GA 30332, USA
}

\date{\today}

\begin{abstract}
    Radiation transport is an essential process in many physical simulations, including cosmological codes which track stellar and diffuse radiation.
    Many present codes utilize ray tracing or moment methods, but these grow increasingly difficult and expensive for large simulations.
    Quantum lattice Boltzmann methods exponentially reduce complexity by propagating radiation over a lattice, allowing for better spatial resolution without needing to introduce adaptive ray tracing or other optimizations.
    However, existing formulations result in poor angular resolution for radiation point sources.
    In this work, we propose an angular redistribution methodology which significantly improves the resolution with minimal increase in complexity.
    We verify the correctness of the method on both classical processors and quantum simulators and outline how the quantum algorithm can be integrated into existing codes.
    This method is an important step towards fully-quantum radiation hydrodynamics simulation which could exponentially accelerate existing cosmological codes, enabling simulations at scales currently infeasible.
\end{abstract}


\maketitle

\section{\label{sec:intro} Introduction}

Computational cosmology, the use of simulations to study the universe as a whole, is an increasingly-important area of physics, as observations have recently become more detailed and numerous with the launch of the {\it James Webb Space Telescope} \cite{Adams2024}. Complex astrophysical systems are often best modeled numerically with the multitude of physical processes and length and time scales involved in cosmic systems.

Many simulations have been performed with extremely large domains, such as the following:
\begin{itemize}
    \item The Renaissance Simulations with comoving volumes of (40 Mpc)$^3$ \cite{OShea2015} performed with the Enzo code \cite{Bryan2014,BrummelSmith2019}.
    \item The Cosmic Dawn (CoDa) III simulations with comoving volumes of (94.4 Mpc)$^3$ \cite{Lewis2022} performed with the RAMSES-CUDATON code \cite{Teyssier2002,Aubert2008}.
    \item The THESAN simulations with comoving volumes of (95.5 Mpc)$^3$ \cite{Kannan2021,Garaldi2022,Smith2022} performed with the AREPO-RT code \cite{Kannan2019}.
\end{itemize}
These require access to large allocations of memory, often demanding additional memory allocation over the course of the simulation and large quantities of auxiliary data. This even limits the capabilities of many simulations, as it can be too computationally expensive to track all of the quantities of interest or physical processes over the whole domain, forcing one to limit their study to a subset of these quantities and processes without becoming memory- or compute-limited.

In such simulations, one of the most important but costliest processes to model is radiative transfer, the propagation of photons from luminous sources, such as stars and active galactic nuclei that are effectively isotropic point sources on cosmological scales, and diffuse gas that emit radiation through radiative processes.
Ray tracing, a commonly-used method, tracks a sample of rays each with an associated flux emitted by every source, growing in cost with both the number of sources and the simulation grid size as one must sufficiently sample the entire medium to obtain accurate results \cite{Wise2011}. Many ray tracing solvers either apply short characteristics, where quantities are interpolated along short paths from the source to the destination \cite{Raga1999,Whalen2006,Mellema2006}, or long characteristics, where quantities are interpolated until the end of the domain \cite{Abel1999,Finlator2009}.
Flux-limited diffusion methods, such as those presented by Ref.~\cite{Turner2001} and Ref.~\cite{Reynolds2009}, utilize a diffusion term tuned to ensure that radiation does not propagate with infinite flux through the simulation grid; these methods are fast but not always accurate.
A wide class of methods known as moment methods solve the radiative transfer equations in terms of their moments, such as radiative energy and flux \cite{Gonzalez2007,Aubert2008,Rosdahl2015,Hopkins2018}. This requires ``closure'' of the moments using the radiation pressure, which is typically calculated in terms of the Eddington tensor.
Hybrid methods also exist, such as variable Eddington tensor (VET) methods which use short characteristics to calculate the radiation tensor but use moment methods to evolve the radiation field \cite{Stone1992,Gnedin2001,Hayes2003,Davis2012}.

\begin{figure*}[ht!]
    \centering
    \subfigure{
        \begin{tikzpicture}[scale=2,baseline=(current bounding box.center)]
            \node at (0.5,1.2) {(a) D2Q4};
            \draw[step=1.0,black,thick] (0.0,0.0) grid (1.0,1.0);
            \draw [ultra thick] [<->] (0.5,0.00) -- (0.5,1.00);
            \draw [ultra thick] [<->] (0.00,0.5) -- (1.00,0.5);
        \end{tikzpicture}
    }
    \hspace{2cm}
    \subfigure{
        \begin{tikzpicture}[scale=2,baseline=(current bounding box.center)]
            \node at (0.5,1.2) {(b) D2Q8};
            \draw[step=1.0,black,thick] (0.0,0.0) grid (1.0,1.0);
            \draw [ultra thick] [<->] (0.01,0.01) -- (0.99,0.99);
            \draw [ultra thick] [<->] (0.01,0.99) -- (0.99,0.01);
            \draw [ultra thick] [<->] (0.5,0.00) -- (0.5,1.00);
            \draw [ultra thick] [<->] (0.00,0.5) -- (1.00,0.5);
        \end{tikzpicture}
    }
    \vspace{1mm}
    \\
    \vspace{1mm}
    \subfigure{
        \begin{tikzpicture}[scale=2,baseline=(current bounding box.center)]
            \node at (1.5,3.2) {(c) D2Q16};
            \draw[step=1.0,black,thick] (1.0,1.0) grid (2.0,2.0);
            \draw[step=1.0,black,thick] (0.0,0.0) grid (3.0,3.0);
            \draw [ultra thick] [<->] (1+0.00,1+0.50) -- (1+1.00,1+0.50);
            \draw [ultra thick] [<->] (1+0.50,1+0.00) -- (1+0.50,1+1.00);
            \draw [ultra thick] [<->] (1+0.00,1+0.00) -- (1+1.00,1+1.00);
            \draw [ultra thick] [<->] (1+0.00,1+1.00) -- (1+1.00,1+0.00);
            \draw [ultra thick] [<->] (1+0.00,1+1.00) -- (1+1.00,1+0.00);
            \draw [ultra thick] [<->] (0.00,1.00) -- (3.00,2.00);
            \draw [ultra thick] [<->] (0.00,2.00) -- (3.00,1.00);
            \draw [ultra thick] [<->] (1.00,0.00) -- (2.00,3.00);
            \draw [ultra thick] [<->] (2.00,0.00) -- (1.00,3.00);
        \end{tikzpicture}
    }
    \hspace{2cm}
    \subfigure{
        \begin{tikzpicture}[scale=2,baseline=(current bounding box.center)]
            \node at (0.0,1.6) {(c) D3Q27};
            \draw [ultra thick] [->] (0,0,0) -- (xyz cs:x=+1);
            \draw [ultra thick] [->] (0,0,0) -- (xyz cs:x=-1);
            \draw [ultra thick] [->] (0,0,0) -- (xyz cs:y=+1);
            \draw [ultra thick] [->] (0,0,0) -- (xyz cs:y=-1);
            \draw [ultra thick] [->] (0,0,0) -- (xyz cs:z=+1);
            \draw [ultra thick] [->] (0,0,0) -- (xyz cs:z=-1);
            \draw [ultra thick] [->] (0,0,0) -- (xyz cs:x=+1.0,y=+1.0);
            \draw [ultra thick] [->] (0,0,0) -- (xyz cs:x=-1.0,y=+1.0);
            \draw [ultra thick] [->] (0,0,0) -- (xyz cs:x=+1.0,y=-1.0);
            \draw [ultra thick] [->] (0,0,0) -- (xyz cs:x=-1.0,y=-1.0);
            \draw [ultra thick] [->] (0,0,0) -- (xyz cs:x=+1.0,z=+1.0);
            \draw [ultra thick] [->] (0,0,0) -- (xyz cs:x=+1.0,z=-1.0);
            \draw [ultra thick] [->] (0,0,0) -- (xyz cs:y=+1.0,z=+1.0);
            \draw [ultra thick] [->] (0,0,0) -- (xyz cs:y=+1.0,z=-1.0);
            \draw [ultra thick] [->] (0,0,0) -- (xyz cs:x=-1.0,z=+1.0);
            \draw [ultra thick] [->] (0,0,0) -- (xyz cs:x=-1.0,z=-1.0);
            \draw [ultra thick] [->] (0,0,0) -- (xyz cs:y=-1.0,z=+1.0);
            \draw [ultra thick] [->] (0,0,0) -- (xyz cs:y=-1.0,z=-1.0);
            \draw [ultra thick] [->] (0,0,0) -- (xyz cs:x=+1.0,y=+1.0,z=+1.0);
            \draw [ultra thick] [->] (0,0,0) -- (xyz cs:x=-1.0,y=+1.0,z=+1.0);
            \draw [ultra thick] [->] (0,0,0) -- (xyz cs:x=+1.0,y=-1.0,z=+1.0);
            \draw [ultra thick] [->] (0,0,0) -- (xyz cs:x=-1.0,y=-1.0,z=+1.0);
            \draw [ultra thick] [->] (0,0,0) -- (xyz cs:x=+1.0,y=-1.0,z=-1.0);
            \draw [ultra thick] [->] (0,0,0) -- (xyz cs:x=-1.0,y=-1.0,z=-1.0);
            \draw [ultra thick] [->] (0,0,0) -- (xyz cs:x=+1.0,y=+1.0,z=-1.0);
            \draw [ultra thick] [->] (0,0,0) -- (xyz cs:x=-1.0,y=+1.0,z=-1.0);
            \draw [thick] [-] (-1,-1,-1) -- (+1,-1,-1);
            \draw [thick] [-] (-1,-1,-1) -- (-1,+1,-1);
            \draw [thick] [-] (-1,-1,-1) -- (-1,-1,+1);
            \draw [thick] [-] (+1,+1,+1) -- (-1,+1,+1);
            \draw [thick] [-] (+1,+1,+1) -- (+1,-1,+1);
            \draw [thick] [-] (+1,+1,+1) -- (+1,+1,-1);
            \draw [thick] [-] (-1,-1,+1) -- (-1,+1,+1);
            \draw [thick] [-] (-1,+1,-1) -- (-1,+1,+1);
            \draw [thick] [-] (-1,-1,+1) -- (-1,+1,+1);
            \draw [thick] [-] (+1,+1,-1) -- (+1,-1,-1);
            \draw [thick] [-] (+1,-1,-1) -- (+1,-1,+1);
            \draw [thick] [-] (+1,+1,-1) -- (-1,+1,-1);
            \draw [thick] [-] (-1,-1,+1) -- (+1,-1,+1);
            \draw [dashed] [-] (-1,-1,-0) -- (+1,-1,-0);
            \draw [dashed] [-] (-0,-1,-1) -- (-0,+1,-1);
            \draw [dashed] [-] (-0,-1,-1) -- (-0,-1,+1);
            \draw [dashed] [-] (+1,+1,+0) -- (-1,+1,+0);
            \draw [dashed] [-] (+0,+1,+1) -- (+0,-1,+1);
            \draw [dashed] [-] (+1,+0,+1) -- (+1,+0,-1);
            \draw [dashed] [-] (-1,+0,+1) -- (-1,+0,-1);
            \draw [dashed] [-] (+0,+1,+1) -- (-0,+1,+1);
            \draw [dashed] [-] (+0,-1,-1) -- (-0,-1,-1);
            \draw [dashed] [-] (+1,+0,+1) -- (-1,+0,+1);
            \draw [dashed] [-] (+1,+0,-1) -- (-1,+0,-1);
            \draw [dashed] [-] (+0,+1,+1) -- (-0,+1,-1);
        \end{tikzpicture}
    }
    \caption{
        Example D$n$Q$m$ lattice Boltzmann schemes in $n$ dimensions and $m$ angular discretizations, showing examples for (a) D2Q4, (b) D2Q8, (c) D2Q16, and (d) D3Q27 cases. It is worth noting that, although not all vectors have the same length, they are equally-distributed in angle so that there is no inherent anisotropy in the system.
    }
    \label{fig:lattice_schemes}
\end{figure*}
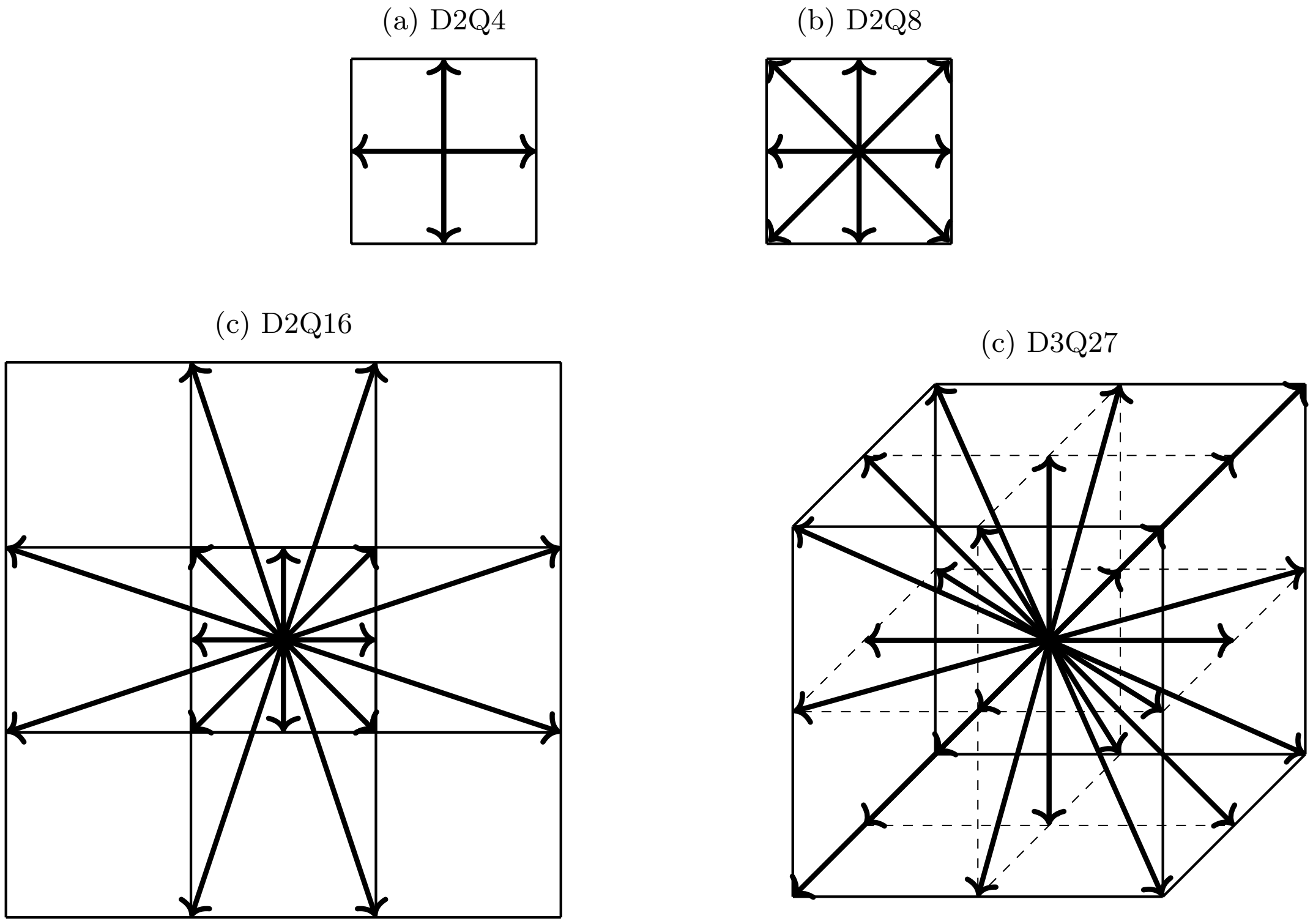

Lattice Boltzmann methods (LBMs), adapted from Lattice Gas Automata methods \cite{McNamaraZanetti1988}, track the radiation intensity over a Cartesian grid of lattice sites along a few discrete directions, as depicted in Figure \ref{fig:lattice_schemes}.
This approach has the advantage that spatial and angular resolution always come at the same cost, regardless of how active or complex a region may be. These can be implemented with a variety of grid geometries as well as with adaptive methods to further optimize the cost.

LBMs have have already been applied to computational fluid dynamics \cite{McNamaraZanetti1988}, heat transfer \cite{Asinari2010}, neutron transport \cite{McCulloch2015, BindraPatil2012}, and even couplings between these \cite{McCulloch2016, BindraPatil2012}. Neutron transport is a particularly similar problem to radiative transfer because its governing equation contains similar contributions, with the primary difference being that the neutron transport problem includes interactions between neutrons and transitions such as nuclear fission. Most importantly, neutron transport applications feature scattering, making such implementations unsuitable for cosmological problems.

Quantum algorithms, which perform computations by manipulating information stored in quantum systems, have been proposed for accelerating classical computations which are known to be computationally-expensive, such as prime factorization, unstructured search, and Hamiltonian simulation. Quantum algorithms have already been considered for their applicability to astrophysical and cosmological simulations. Most works focus on variational quantum algorithms, which encode an objective function into a Hamiltonian and iteratively search for a statevector which optimizes the objective value, representing the most likely state of the system \cite{Mocz2021,Kaufman2019,Joseph2021,Joseph2022,Wang2024}. There has also been some work into Hamiltonian evolution via Trotterization, where a system Hamiltonian is derived and used to compute discrete unitary time evolution operators; these time evolution operators can be applied to a statevector which encodes the initial conditions in order to obtain the time-evolved statevector \cite{Joseph2021,Liu2021}.

The radiative transfer problem in particular has also been studied by writing quantum adaptations of classical methods such as ray tracing \cite{Lu2022_1, Lu2022_2, Lu2023, Mosier2023, Santos2025} and random walks \cite{Lee2025}, as well as through novel quantum algorithms for differential equations \cite{Gaitan2024}. Quantum LBMs take advantage of the ability to easily parallelize computations and store information in a more scalable manner in quantum computing, exponentially reducing the cost associated with high-resolution simulations even before performing optimizations. They were first demonstrated by the authors of Ref.~\cite{Yepez2001} for lattice gas automata, following similar work in Ref.~\cite{Meyer1996} for quantum cellular automata applied to gases. Many works have studied further studied quantum LBMs for fluid dynamics since, including Refs. \cite{Budinski2021,Kocherla2024,Lee2024,Wawrzyniak2025_1,Wawrzyniak2025_2,Yamazaki2025}. The reader may reference Refs. \cite{Love2019,Succi2023} for a broader review of quantum methods for fluid and fluid-like simulations.

The authors of Ref.~\cite{Igarashi2024}, in particular, formulated the first quantum algorithm to simulate radiative transfer in one dimension and also discuss extensions into two dimensions, but their formulation is only applicable in scattering media with linear or planar geometries. Despite the progress of these existing works, there is no demonstration of quantum lattice Boltzmann radiative transfer in two or three dimensions. Although such methods can be extended with high accuracy to higher dimensions in scattering media \cite{Igarashi2024}, the methodology is not accurate for point sources in non-scattering media, as we demonstrate in this work.

This work demonstrates a lattice Boltzmann methodology which utilizes angular redistribution to improve simulation accuracy for cosmological radiation transport problems. In section \ref{sec:lbms}, we provide background on lattice Boltzmann methods, particularly the one that is the basis of our scheme, and compare them to other methods.
In section \ref{sec:alg}, we provide an overview of the quantum algorithm implemented in this work, followed by the motivation behind and formulation of our novel angular redistribution step.
In section \ref{sec:impl}, we detail how the quantum algorithm is implemented and executed, including optimizations in how the quantum circuit is constructed classically.
In section \ref{sec:results}, we present results for a variety of test simulations, demonstrating the algorithm's accuracy and effectiveness. We conclude in section \ref{sec:conclusion} with a discussion of further steps which will be taken to improve upon this work.

\section{\label{sec:lbms} Lattice Boltzmann methods}

\subsection{Overview}

LBMs are constructed by fixing the number of grid dimensions $n$ and then selecting, at each grid point, a number of discrete angular directions $m$ along which a quantity (here, the radiation intensity) can propagate. This results in what is often referred to as a D$n$Q$m$ scheme, where the variables $n$ and $m$ specifically identify the scheme. LBMs for radiative transfer have already been implemented classically with high-accuracy with unstructured grids \cite{Mondal2008, Mondal2009, Wang2017, Liu2020, Ymeli2023, Wu2024, Liu2024}. Moreover, $m$ need not be a constant across grid cells or over time. However, because the implementation of LBMs requires a complete specification of where quantities travelling along any of the discrete directions should be propagated, such adaptive or unstructured approaches require more careful formulation, such as to ensure conservation of flux. This aspect is especially the case for quantum algorithms, as adaptive schemes would require real-time construction and compilation of the algorithm circuits. In this work, we only consider constant lattice Boltzmann schemes. For a constant scheme, every iteration is performed using the same algorithm, possibly excluding state preparation of the source if it changes.

Once the number of discrete angular directions has been selected, one must compute each of the direction vectors. For $m$ equally-spaced angular directions on a plane in two dimensions, the angles are exactly fractions of unity:
\begin{align}
    \vec\theta = \left\{\dfrac{2\pi\mu}{m} : \mu \in 0, 1, \ldots, m\right\}.
\end{align}
Then, their vectors can be expressed as
\begin{align}
    \vec c_\mu = \left(\cos{\dfrac{2\pi\mu}{m}}, \sin{\dfrac{2\pi\mu}{m}}\right).
\end{align}
Typically for LBMs, these vectors are multiplied by constants such that all vectors lie along integer spacings on a Cartesian grid, as depicted in Figure \ref{fig:lattice_schemes}. On cubes in three dimensions, one adds an additional component due to the vertical offset.

\subsection{Comparison with other methods}
One may view lattice Boltzmann schemes as a modification of traditional ray tracing but with all angles adjusted to fit one of finitely-many angular directions. Although this adjustment poses the challenge of resolving a lower angular resolution, the deterministic structure makes encoding less expensive because one does not need to track the exact angle at which a ray is travelling, only the index of its angle in a finite list. In the present setting of noisy intermediate-scale quantum (NISQ) devices, the limited number of qubits sets a stricter constraint than, for comparison, the maximum circuit depth at which quantum computation can be coherently performed; so, this reduction in memory makes the algorithm a more viable candidate for NISQ applications. This constraint on space complexity is further tightened by the need for quantum error correction, which multiplies the qubit count requirement.

Also, while ray tracing methods track the radiation intensity at the cell vertices or center, LBMs track it on more points per lattice site, allowing for better angular resolution at each grid point.
This increase in the number of points increases the classical memory cost linearly, but it only increases the quantum cost logarithmically, making this approach more affordable for quantum simulations.
Some classical methods besides LBMs also discretize angles, such as the short-characteristics method implemented by the authors of Ref.~\cite{Davis2012}. The form of LBMs we use in this work are, in some ways, simpler than short-characteristics methods, particularly in the fact that we do not perform interpolation as a central step in our algorithm, though it can still be applied. In fact, the discretization of LBMs is closely related to the method of characteristics \cite{Kruger2017}. The lattice direction vectors are analogous to the characteristics of integration.

Due to the propagation of radiation according to finite angular direction vectors, LBMs treat the speed of light as a finite quantity. This is contrast with some approaches which use an ``infinite'' speed of light, such as some cases of ray tracing. In these other approaches, photons are transported as far as possible during each radiative transfer timestep. This requires more computation per timestep and notably limits the accuracy of simulations involving relativistic effects. Because LBMs treat the speed of light as a finite quantity, our approach is favorable for simulations involving relativistic effects, such as neutron stars and black holes. We leave their study to future works and only consider the non-relativistic equations in this work.

\section{\label{sec:alg} General quantum algorithm}

The reformulation of a classical algorithm as a quantum algorithm requires the representation of each computational step as a unitary linear operator, as these are the only physical transformations which can be performed on a quantum mechanical system. Due to the fixed, structured nature of lattice Boltzmann methods, it is more straightforward to derive from it a quantum algorithm.

The general cosmological radiative transfer equation in non-scattering media is
$$\dfrac{1}{c} \DD[I_f]{t} + \dfrac{\vec \Omega \cdot \nabla I_f}{\overline{a}} - \dfrac{H}{c} \left(f \DD[I_f]{f} - 3I_f\right) = -\kappa_f I_f + S_f,$$
where $c$ is the speed of light, $I_f$ is the radiation intensity of a specific frequency $f$, $\vec \Omega$ is the angular direction vector, $\overline{a} = a/a_\mathrm{em}$ is the ratio of scale factors at the current time and time of emission, $H = \dot a/a$ is the Hubble parameter, $\kappa_f$ is the absorption coefficient at frequency $f$, and $S_f$ is the source term at frequency $f$ \cite{Wise2011}. While this is similar to the Boltzmann transport equation, it notably lacks collision and scattering terms. Collision effects in the traditional Boltzmann sense are non-existent for radiation transport problems, though scattering can be included if it is relevant to a specific problem. Our formulation of the quantum lattice Boltzmann algorithm for this problem includes a step for scattering for the case where it is necessary.

For problems where cosmological redshifting and dilution are negligible, this expression reduces to the local approximation \cite{Wise2011}:
$$\DD[I]{t} + \vec \Omega \cdot \nabla I = -\kappa I + S,$$
where we also now take $c = 1$.

For traditional LBMs in any number of dimensions with $m$ angular directions, we may discretize this expression to obtain the following form of the radiative transfer equation along a specific direction indexed by $1 \leq \mu \leq m$:
$$I_\mu(x + c_\mu, t + \delta_t) - I_\mu(x, t) = -\kappa \delta_t I_\mu + \dfrac1m \delta_t S_\mu(x).$$

\begin{figure*}[ht!]
    \begin{center}
        \begin{quantikz}
            \lstick{$\ket{L}^{\otimes n_l}$}
                & \gate[5]{U_\mathrm{SP}}
                & \gate[5]{U_\mathrm{AS}}
                & \gate[5]{U_\mathrm{AE}}
                & \gate[5]{U_\mathrm{P}}
                & \gate[5]{U_\mathrm{AR}}
                & \meter{}
                &
            \\
            \lstick{$\ket{D}^{\otimes n_d}$}
                &
                &
                &
                &
                &
                & \meter{}
                &
            \\
            \lstick{$\ket{S}$}
                &
                &
                &
                &
                &
                & \meter{}
                &
            \\
            \lstick{$\ket{M}$}
                &
                &
                &
                &
                &
                & \meter{}
                &
            \\
            \lstick{$\ket{A}^{\otimes n_a}$}
                &
                &
                &
                &
                &
                & \meter{}
                &
            \\
        \end{quantikz}
    \end{center}

    \caption{The full quantum circuit for LBMs including our angular redistribution step. The circuit contains registers indicating the lattice site $\ket{L}^{\otimes n_l}$, the radiation direction $\ket{D}^{\otimes n_d}$, the switch between intensity and source terms $\ket{S}^{\otimes n_s}$, and the extra ancillae qubits $\ket{A}^{\otimes n_a}$. The circuit begins with state preparation with $U_{SP}$, then absorption and scattering with $U_{AS}$, then absorption and emission with $U_{AE}$, then propagation with $U_P$, then angular redistribution with $U_{AR}$, and ends with final measurements.}
    
    \label{fig:full_circuit}
\end{figure*}
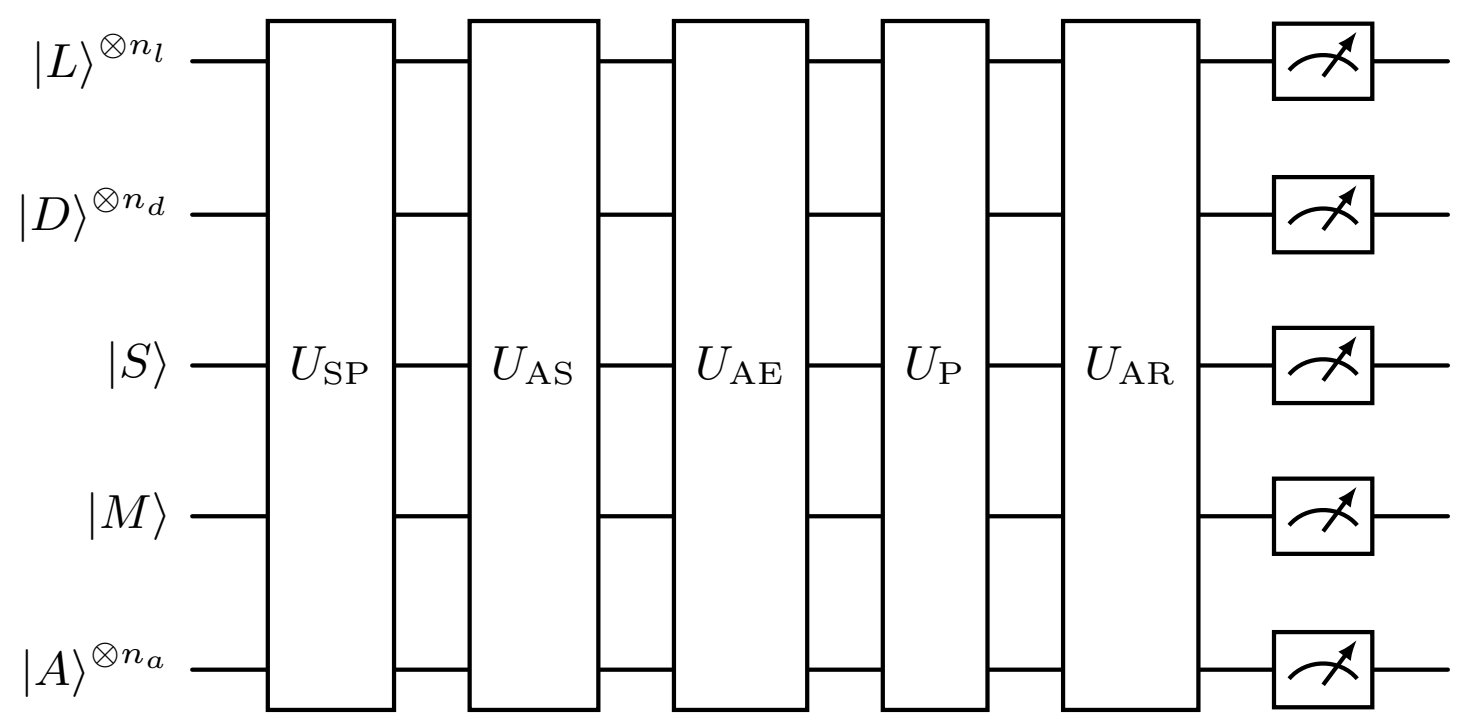

The full quantum circuit is shown in Figure \ref{fig:full_circuit}. We order the steps in a way which allows for the behavior we desire during simulations. In particular,
\begin{enumerate}
    \item Angular redistribution is performed after absorption and emission so that emitted radiation is redistributed during the same step;
    \item Boundary conditions are applied during measurement to reduce quantum gate cost.
\end{enumerate}

\subsection{\label{sec:encoding} State Preparation}
For a single-mode radiation field in three dimensions, the quantum statevector for the initial conditions is
\begin{widetext}
$$\ket{\phi} = \ket{0}^{\otimes5}_a \dfrac{1}{\norm{\phi}} \sum_{(i, j,k)=(0,0,0)}^{\left(L_X-1, L_Y-1, L_Z-1\right)} \sum_{\mu=0}^{m-1} \Bigg(I^{(i, j, k)}_\mu \ket{M}_m \ket{0}_s \ket{\mu}_d \ket{i, j, k}^{\otimes \log_2{m}}_l + \dfrac12 \delta_t S^{(i, j, k)}_\mu \ket{0}_m \ket{1}_s \ket{\mu}_d \ket{i, j, k}^{\otimes \log_2{m}}_l\Bigg).$$
\end{widetext}

Here, the subscript $a$ indicates the ancilla register, the subscript $m$ indicates the angular redistribution mask register, the subscript $s$ indicates the switch register, the subscript $d$ indicates the direction register, and the subscript $l$ indicates the lattice data register. The indices $i,j,k$ indicate the lattice point at coordinate $(i,j,k)$, ranging from $(0,0,0)$ to $(L_X,L_Y,L_Z)$.

The lattice data qubits indicate the position of any individual vector in the quantum state as binary indices.
For a simulation grid with $M_0$ total lattice sites (where separate directions are not considered separate sites), the number of lattice data qubits is $n_l = \lceil{\log_2{M_0}\rceil}$. Because of the ceiling function, it is most efficient, if possible, to work with grid sizes which are powers of 2.
In this paper, we focus primarily on the three-dimensional formulation, but fewer dimensions can be handled similarly.

The direction qubits indicate the angular direction of any vector in the quantum state as binary indices.
A scheme with $m$ angular directions requires $n_d = \lceil{\log_2{m}\rceil}$ direction qubits. This makes the quantum method particularly advantageous because one can double the number of angular directions at the cost of only an additional qubit.

The switch qubit allows us to ``switch'' between acting on intensity terms and source terms by controlling any action on its state.
Since there are only two such modes of operation, a single switch qubit is sufficient for any LBM scheme, independent of the number of dimensions or angular directions; that is, $n_s = 1$.

The multiple ancillary qubits are necessary for various steps in the algorithm, such as performing controlled operations. We describe their individual uses in more detail when describing the algorithm.
For any simulation scheme and size, two ancilla qubits are needed for absorption and emission. For a D$n$Q$m$ scheme, $\lceil\log_2{m}\rceil$ ancilla qubits are needed for propagation. Absorption and scattering requires an additional 2 ancilla qubits; angular redistribution requires an additional 2 ancilla qubits in 2D and 3 in 3D.
For any simulation in 2D, the necessary number of ancilla qubits is $n_a = 5$.
Although not discussed in further detail in this work, a reduction in the ancilla qubit requirement is one area for improvement of our quantum LBM formulation. We do make one optimization in our implementation; if a certain step (such as scattering or emission) is not used in a simulation, then any associated ancillae are not allocated.

The method used to prepare this initial state is important because the statevector encoding of the initial state of the system is not necessarily the result of a simple quantum circuit and may require many non-trivial and time-consuming entangling operations which reduce the overall fidelity of the quantum algorithm.

\subsection{\label{sec:absorption_scattering} Absorption and Scattering}

We begin by considering the absorption and scattering of the radiation. In the non-scattering form of the radiative transfer equation most commonly used in cosmological problems, this step is not necessary, but we describe it here for the sake of generality.

We can perform more advanced boundary conditions during the post-selection step, detailed in section \ref{sec:measurement} Otherwise, we assume that the absorption coefficient is constant at all lattice sites. Continuously-varying absorption coefficients may be implemented naively by controlling the quantum circuit described in this section on sets of lattice sites with the same absorption coefficients and using different values of the absorption coefficient at each site.

The absorption and scattering processes are captured in the following equation:
$$I_\mu(x, t) = \left(1 - \kappa \delta_t\right) I_\mu(x, t) + \dfrac{1}{m} \sum_{\nu=0, \nu\neq\mu}^{m-1} \sigma_{\mu\nu} I_\nu(x, t)$$

This can be expressed by the following operator:
\begin{align*}
    \hat U_{AS} = (1-\kappa\delta_t) \I_{amsdl} + \dfrac{1}{m} \sigma \delta_t \left(\I_d + \X_d\right) \otimes \I_{amsl},
\end{align*}
where $\I_{amsdl}$ is the identity operator over all of the registers, $\I_d$ is the identity operator over the direction register, $\X_d$ is the Pauli-$X$ operator over the direction register, and $\I_{asl}$ is the identity operator over all of the registers excluding the direction register. In this definition, the first term captures globally-constant absorption with coefficient $0 \leq \kappa \leq 1$ and the second term captures globally-constant isotropic scattering with coefficient $0 \leq \sigma \leq 1$. The circuit for this step is shown in Figure \ref{fig:absorption_emission_circuit}.

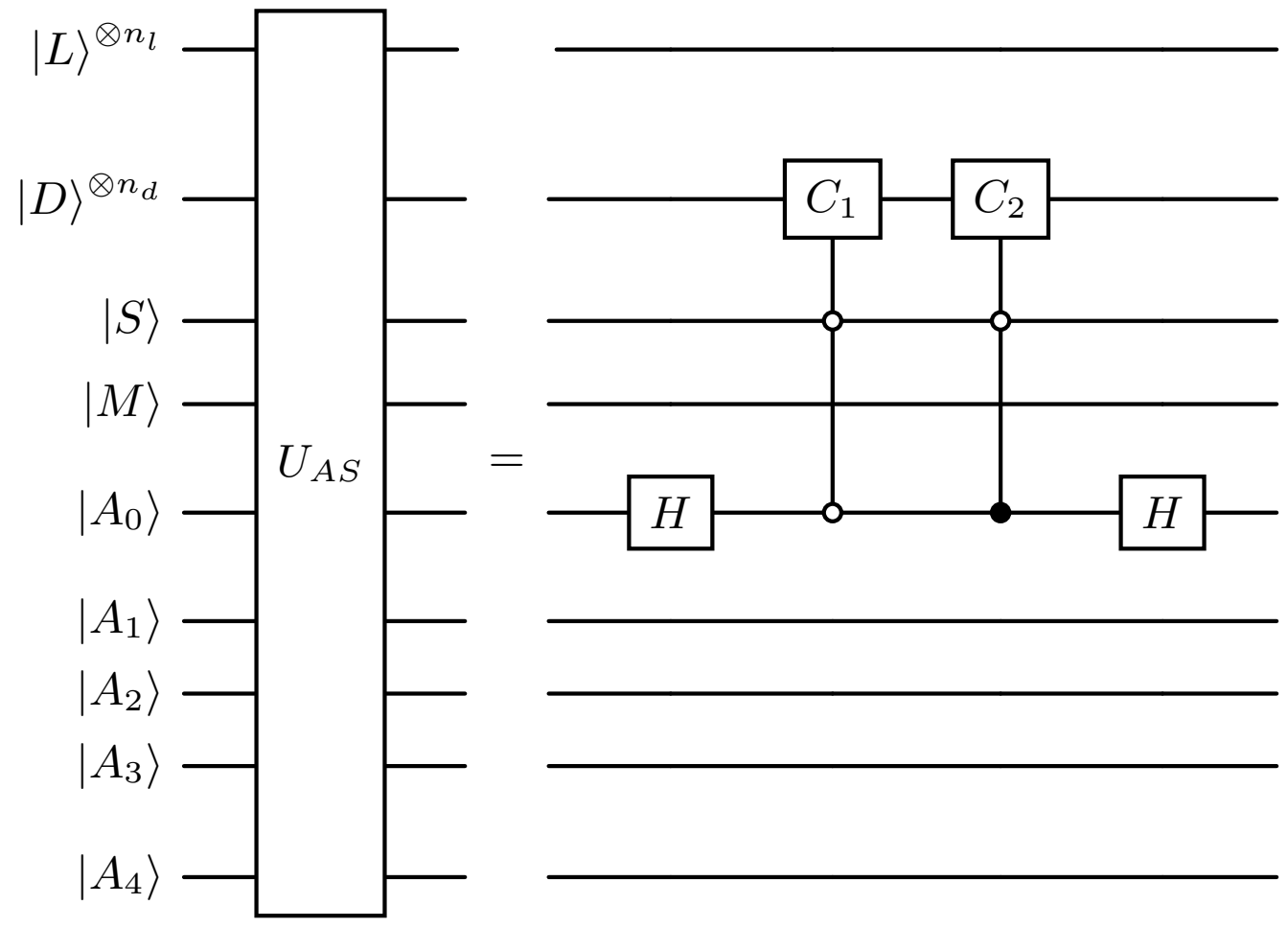
\begin{figure*}[ht!]
\centering
    \begin{center}
        \begin{quantikz}
            \lstick{$\ket{L}^{\otimes n_l}$}
                & \gate[9]{U_{AS}}
                & \midstick[9,brackets=none]{=}
                &
                &
                &
                &
                &
            \\
            \lstick{$\ket{D}^{\otimes n_d}$}
                &
                &
                & 
                & \gate{C_1}
                & \gate{C_2}
                & 
                &
            \\
            \lstick{$\ket{S}$}
                &
                &
                & 
                & \octrl{0} \wire[u]{q}
                & \octrl{0} \wire[u]{q}
                & 
                & 
            \\
            \lstick{$\ket{M}$}
                &
                &
                & 
                & \wire[u]{q}
                & \wire[u]{q}
                & 
                & 
            \\
            \lstick{$\ket{A_0}$}
                &
                &
                & \gate{H}
                & \octrl{0} \wire[u]{q}
                & \ctrl{0} \wire[u]{q}
                & \gate{H}
                &
            \\
            \lstick{$\ket{A_1}$}
                &
                &
                & 
                & 
                & 
                & 
                &
            \\
            \lstick{$\ket{A_2}$}
                &
                &
                & 
                & 
                & 
                & 
                &
            \\
            \lstick{$\ket{A_3}$}
                &
                &
                & 
                & 
                & 
                & 
                & 
            \\
            \lstick{$\ket{A_4}$}
                &
                &
                & 
                & 
                & 
                & 
                & 
            \\
        \end{quantikz}
    \end{center}
    \caption{The quantum circuit for the absorption and scattering for arbitrary lattice schemes. In non-scattering media, this circuit is equivalent to the identity and thus can be skipped. In this and the following quantum circuit diagrams, we use the convention where an open circle corresponds to conditioning on the 0 state (so that the controlled operator is only applied if that qubit is in the 0 state) and a closed circle corresponds to conditioning on the 1 state (so that the controlled operator is only applied if that qubit is in the 1 state).}
    \label{fig:absorption_scattering_circuit}
\end{figure*}

Note that, in non-scattering media, $\sigma_\mu = 0$ for all $\mu$, in which case this step reduces to a constant multiple of the identity operator. However, such an operator is non-unitary, so it is not a valid quantum transformation; we instead handle this absorption effect by tracking the absorption factor $1-\kappa\delta_t$ in software. We follow this general procedure at each step of the simulation: we implement the unitary (and thus norm-preserving) dynamics of our algorithm as quantum circuits and keep track of global factors not equal to unity in software. The statevector, therefore, only stores the relative intensities, and the cumulative product of the factors stores the overall scale of the intensity over the entire grid.

\subsection{\label{sec:absorption_emission} Absorption and Emission}

In order to emit the radiation, we must add the amplitudes of the source terms to the corresponding amplitudes of the intensity terms at each lattice site.
This corresponds to the following update equation:
$$I_\mu(x, t + \delta_t) = (1 - \kappa \delta_t) I_\mu(x, t) + \dfrac{1}{m} \delta_t S_\mu(x, t)$$

This can be expressed by the following operator:
\begin{align*}
    \hat U_{AE} = \I_{amdl} \otimes \left(\I_{s} + \dfrac12 \X_{s} + \dfrac12 \Z_{s} \X_{s}\right).
\end{align*}

\begin{figure*}[ht]
\centering
    \begin{center}
        \begin{quantikz}
            \lstick{$\ket{L}^{\otimes n_l}$}
                & \gate[9]{U_{AE}}
                & \midstick[9,brackets=none]{=}
                &
                &
                &
                &
                &
                &
                &
                &
            \\
            \lstick{$\ket{D}^{\otimes n_d}$}
                &
                &
                & 
                & 
                & 
                & 
                & 
                & 
                & 
                &
            \\
            \lstick{$\ket{S}$}
                &
                &
                & 
                & \gate{I}
                & \gate{I}
                & \gate{X}
                & \gate{X}
                & \gate{Z}
                & 
                & 
            \\
            \lstick{$\ket{M}$}
                &
                &
                &
                & \wire[u]{q}
                & \wire[u]{q}
                & \wire[u]{q}
                & \wire[u]{q}
                & \wire[u]{q}
                & 
                &
            \\
            \lstick{$\ket{A_0}$}
                &
                &
                &
                & \wire[u]{q}
                & \wire[u]{q}
                & \wire[u]{q}
                & \wire[u]{q}
                & \wire[u]{q}
                & 
                &
            \\
            \lstick{$\ket{A_1}$}
                &
                &
                & \gate{H}
                & \octrl{0} \wire[u]{q}
                & \octrl{0} \wire[u]{q}
                & \ctrl{0} \wire[u]{q}
                & \ctrl{0} \wire[u]{q}
                & \ctrl{0} \wire[u]{q}
                & \gate{H}
                &
            \\
            \lstick{$\ket{A_2}$}
                &
                &
                & \gate{H}
                & \octrl{0} \wire[u]{q}
                & \ctrl{0} \wire[u]{q}
                & \octrl{0} \wire[u]{q}
                & \ctrl{0} \wire[u]{q}
                & \ctrl{0} \wire[u]{q}
                & \gate{H}
                &
            \\
            \lstick{$\ket{A_3}$}
                &
                &
                & 
                & 
                & 
                & 
                & 
                & 
                & 
                & 
            \\
            \lstick{$\ket{A_4}$}
                &
                &
                & 
                & 
                & 
                & 
                & 
                & 
                & 
                & 
            \\
        \end{quantikz}
    \end{center}
    \caption{The quantum circuit for the absorption and emission step for arbitrary lattice schemes. The Hadamard gates put the ancillas in a four-way superposition and the controlled gates correspond to each term in the scattering equation. The double controlled identity gates can be skipped, but we visualize it here for completeness. The use of two controlled identity gates is itself redundant but shows that this operation is weighted twice as much as the $\X$ and $\Z\X$ operations, as seen in the operator definition.}
    \label{fig:absorption_emission_circuit}
\end{figure*}
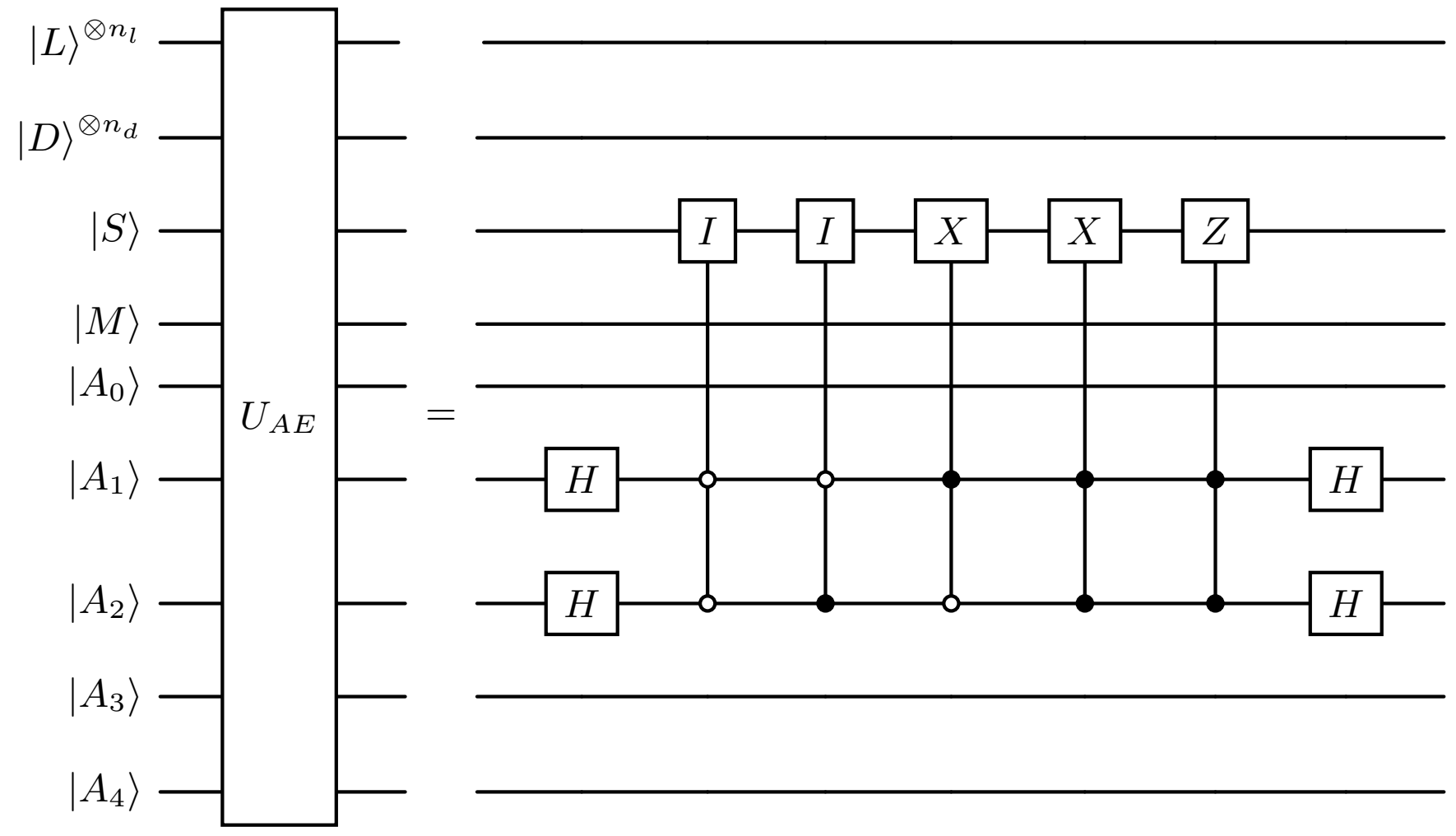

This circuit for this step is shown in Figure \ref{fig:absorption_emission_circuit}. The Hadamard gates put the ancilla qubits in an equal superposition of all possible states. Then, the controlled gates selectively apply the operators from each term of the operator.

\subsection{\label{sec:propagation} Propagation}

Once the local radiative effects have been simulated, one must handle the propagation of radiation by streaming the distribution along each angular direction. The initial challenge in formulating the propagation operators comes in identifying the appropriate boundary conditions to use, so that radiation at the edges of the simulation domain is appropriately handled. In this work, we treat propagation as periodic during this step and apply the true boundary conditions during post-selection. In the LBM formulation of radiative transfer, the propagation step is expressed in the following form:
$$I_\mu(x + c_\mu, t + \delta_t) = I_\mu(x, t).$$

To formulate this as a quantum operator, first let us define the quantum walk operator:
\begin{align*}
    \hat V_{P_\mu} &= \sum_{(i,j,k)=(0,0,0)}^{(L-1,M-1,N-1)} \ket{\vec x+\vec c_\mu}_l \bra{\vec x}_l.
\end{align*}
This operator can be interpreted as shifting all positions $\vec x$ by a velocity vector $\vec c_\mu$ to $\vec x + \vec c_\mu$. Then, propagation along the $\mu$ angular direction can be expressed by the following operator:
\begin{align*}
    \hat U_{P_{\mu}}
        &= \I_{am} \otimes \ket{0}_s \bra{0}_s \otimes \ket{\mu}_d \bra{\mu}_d \otimes \hat V_{P_\mu} \\
        &\ + \I_{am} \otimes \ket{0}_s \bra{0}_s \otimes \ket{\mu^\perp}_d \bra{\mu^\perp}_d \otimes \I_{l} \\
        &\ + \I_{am} \otimes \ket{1}_s \bra{1}_s \otimes \I_{d} \otimes \I_{l}.
\end{align*}
Although some factors in this expression can be regrouped, we have written it this way to emphasize the three distinct ways in which the operator behaves:
\begin{enumerate}
    \item The first term selects intensity terms (using the $\ket{0}_s \bra{0}_s$ projector) along the $\mu$ direction (using the $\ket{\mu}_d \bra{\mu}_d$ projector) and replaces each lattice position $\vec x$ with $\vec{x} + \vec{c}_\mu$ (using each term in the formula for $\hat V_{P_\mu}$).
    \item The second term selects intensity terms (using the $\ket{0}_s \bra{0}_s$ projector) along all directions orthogonal to $\mu$ (using the $\ket{\mu^\perp}_d \bra{\mu^\perp}_d$ projector) and acts trivially on the lattice positions (using the identity).
    \item The third term selects source terms (using the $\ket{1}_s \bra{1}_s$ projector) along every direction (using the identity) and acts trivially on the lattice positions (using the identity).
\end{enumerate}
As one may observe, the non-trivial propagation dynamics are entirely in the first term, while the second and third terms account for cases in which no propagation occurs along a given direction. Geometrically, this equation takes intensity terms with non-zero projection along the $\ket{\mu}_d$ direction and translates their $\ket{\vec x}_l$ component to $\ket{\vec x + \vec c_\mu}_l$.

In the case of periodic boundary conditions, this operation is mathematically equivalent to an independent change of basis of each $\mu$-space, a unitary transformation which we can perform directly as a quantum circuit. This can be seen, for example, by computing $\hat V_{P_\mu}$ for a D1Q2 lattice along a single direction (indicated by $+1$):
\begin{align*}
    \hat V_{P_\mu} &= \sum_{i=0}^{M-1} \ket{i+1 \mod{M}}_l \bra{i}_l \equiv \begin{pmatrix}
        0 & 0 & \cdots & 0 & 1 \\
        1 & 0 & \cdots & 0 & 0 \\
        0 & 1 & \cdots & 0 & 0 \\
        \vdots & \vdots & \ddots & \vdots & \vdots \\
        0 & 0 & \cdots & 1 & 0 \\
    \end{pmatrix},
\end{align*}
in which case we see that the columns of this matrix are the rearranged columns of the identity matrix, indicating that this operator is equivalent to a change of basis.

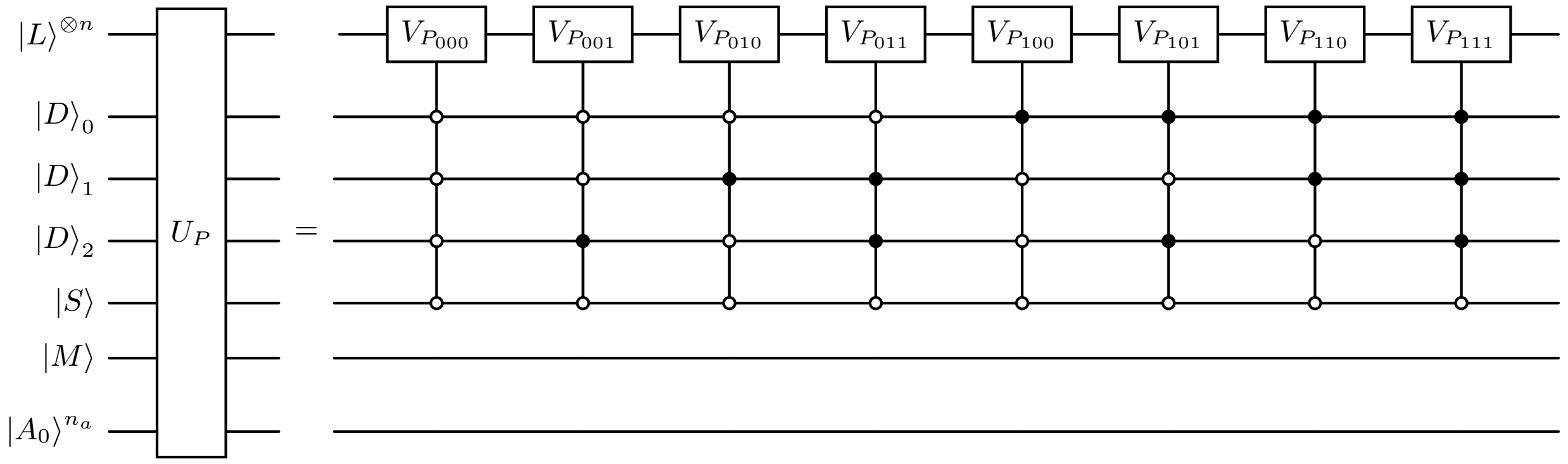
\begin{figure*}[ht!]
    \begin{center}
        \begin{quantikz}
            \lstick{$\ket{L}^{\otimes n}$}
                & \gate[7]{U_{P}}
                & \midstick[7,brackets=none]{=}
                & \gate{V_{P_{000}}}
                & \gate{V_{P_{001}}}
                & \gate{V_{P_{010}}}
                & \gate{V_{P_{011}}}
                & \gate{V_{P_{100}}}
                & \gate{V_{P_{101}}}
                & \gate{V_{P_{110}}}
                & \gate{V_{P_{111}}}
                &
            \\
            \lstick{$\ket{D}_0$}
                &
                & 
                & \octrl{0} \wire[u]{q}
                & \octrl{0} \wire[u]{q}
                & \octrl{0} \wire[u]{q}
                & \octrl{0} \wire[u]{q}
                & \ctrl{0} \wire[u]{q}
                & \ctrl{0} \wire[u]{q}
                & \ctrl{0} \wire[u]{q}
                & \ctrl{0} \wire[u]{q}
                & 
            \\
            \lstick{$\ket{D}_1$}
                &
                & 
                & \octrl{0} \wire[u]{q}
                & \octrl{0} \wire[u]{q}
                & \ctrl{0} \wire[u]{q}
                & \ctrl{0} \wire[u]{q}
                & \octrl{0} \wire[u]{q}
                & \octrl{0} \wire[u]{q}
                & \ctrl{0} \wire[u]{q}
                & \ctrl{0} \wire[u]{q}
                & 
            \\
            \lstick{$\ket{D}_2$}
                &
                & 
                & \octrl{0} \wire[u]{q}
                & \ctrl{0} \wire[u]{q}
                & \octrl{0} \wire[u]{q}
                & \ctrl{0} \wire[u]{q}
                & \octrl{0} \wire[u]{q}
                & \ctrl{0} \wire[u]{q}
                & \octrl{0} \wire[u]{q}
                & \ctrl{0} \wire[u]{q}
                & 
            \\
            \lstick{$\ket{S}$}
                &
                & 
                & \octrl{0} \wire[u]{q}
                & \octrl{0} \wire[u]{q}
                & \octrl{0} \wire[u]{q}
                & \octrl{0} \wire[u]{q}
                & \octrl{0} \wire[u]{q}
                & \octrl{0} \wire[u]{q}
                & \octrl{0} \wire[u]{q}
                & \octrl{0} \wire[u]{q}
                & 
            \\
            \lstick{$\ket{M}$}
                &
                &
                & 
                & 
                & 
                & 
                & 
                &
                & 
                & 
                &
            \\
            \lstick{$\ket{A_0}^{n_a}$}
                &
                &
                & 
                & 
                & 
                & 
                & 
                &
                & 
                & 
                &
            \\
        \end{quantikz}
    \end{center}
    \caption{The quantum circuit for the propagation step for a D2Q8 lattice scheme. Each controlled $V_{P_\mu}$ operator selectively propagates intensities along each of the $8$ lattice directions. These are indexed by their binary representations as this gives a direct labelling of the directions by the direction qubit states.}
    \label{fig:propagation_circuit}
\end{figure*}

The quantum circuit for this step, specifically for a D2Q8 lattice scheme, is shown in Figure \ref{fig:propagation_circuit}. Because we are working with equally-spaced propagation angles and assuming the same flow velocity in all directions, each of the propagation operators is applied in the same way without any reweighting. For each propagation direction, we control the propagation operator for that direction on the intensity states, since they have their switch qubit in the $\ket{0}_s$ state, and on the combination of direction qubit states corresponding to that direction, labelled by an $n_d$-bit string.

It is worth noting that, under periodic boundary conditions on a uniform grid, the propagation unitaries $\hat U_{P_{+\mu}}$ and $\hat U_{P_{-\mu}}$, where $+\mu$ and $-\mu$ correspond to exactly-opposite directions, are controlled gates whose underlying gate actions are Hermitian adjoints | that is, $\hat V_{P_{-\mu}} = \hat V_{P_{+\mu}}^\dagger$.
This symmetry is advantageous because the computation of the propagation unitaries can be the most expensive step in preparing the quantum circuits, and this symmetry allows one to complete this step by only computing the underlying unitaries along half of the directions and taking their adjoints to obtain the remaining unitaries. Taking the adjoint is a much cheaper operation than full reconstruction in this case.

In general, however, we may have to decompose the operation into a linear combination of unitaries, as proposed in Ref.~\cite{Childs2012}. Intuitively, if the desired propagation scheme can result in scenarios where two distinct initial distributions are taken to the same final distribution, then we lose the surjectivity of the propagation step, and thus its unitarity. This would occur, for example, if the absorbing boundary conditions were to be integrated into the propagation step. After attempting such an approach, we found that it results in much more complex unitary decompositions which lack simple decompositions into linear combination. The same concern may arise in certain unstructured or non-uniform grids.

\subsection{\label{sec:angular_redistribution} Angular redistribution}

The other steps in this section completely describe a traditional LBM. However, when applied to non-scattering media in two or more dimensions, notably the setting of most cosmology simulations, such methods leave behind gaps in the intensity field due to the angular discretization. All existing works consider scattering media or only one dimension, cases in which the problems do not appear. Such artifacts indicate that LBMs are insufficient in their traditional forms, and there is a need to resolve this issue before applying it to astrophysical simulations.

We first attempted standard approaches for resolving this issue, namely increasing the number of grid points (an increase in spatial resolution) and number of discrete directions (an increase in angular resolution). We found that increasing the spatial resolution does not reduce the gaps at all, while increasing the angular resolution only reduces the gaps in the immediate vicinity of a source.
After visually inspecting the simulation data in various ways, we made the key observation that the gaps arise because the traditional propagation step does not account for the full area (in 2D) or volume (in 3D) onto which the radiation extends.

Similar effects have been observed in the past for classical ray simulations, known as ray effects in the literature. The authors of \cite{Chai1993} have made a similar argument, but here we rigorously show that improving the resolution alone does not solve the issue. That is, for any traditional LBM scheme applied to non-scattering media, there exists a non-zero angular sector along which the distribution is always zero, independent of the spatial and (finite) angular resolution.

To see this, consider a D$n$Q$m$ scheme with $n \geq 2$ and finite $m$. Since $m$ is finite, we may order each of the discretized angular directions within a single plane. Let $\theta_\mu$ be the angle associated with angular direction $\mu$, where $\mu \in \mathbb{Z}_m$, and $\theta_\mu < \theta_{\mu+1}$ due to our ordering. Then, for each $\mu$, there exists some lattice site at angle $\theta_\mu < \theta_\mu' < \theta_{\mu+1}$, but this angle is not in our set of angular directions along that plane, so any lattice site located along this angle relative to the source will remain empty for the duration of the simulation.

This argument also rules out the possibility of beam tracing methods, such as those first proposed by Ref.~\cite{Heckbert1984}, resolving the gaps, since these effectively only add a finite number of rays around directions which are already angularly-resolved to add ``thickness'' to the rays.

This issue was solved by the authors of Ref.~\cite{Camminady2020} for scattering media, but our issue exists in non-scattering media. The authors of Refs. \cite{Camminady2019,Li2025} solve the problem by rotating their ordinate axes every timestep, but this is both limited in its accuracy (one can only rotate the axes through finitely-many angles during a finite simulation, effectively equating this to increasing the number of angular directions) and extremely expensive for a quantum circuit (it would require recomputing the propagation unitaries for the lattice every rotation).

This issue can also be remedied in post through interpolation, which is done in principle with many ray tracing methods including short-characteristics methods \cite{Raga1999}. However, performing such interpolation can be expensive for large grids, diminishing any computational advantage we may have gained elsewhere. It also raises the question of how one constructs the appropriate multidimensional interpolation stencils, which may be even more difficult when working with unstructured grids. Methods such as those which use short-characteristics also propagate radiation along arbitrary directions from only a subset of the most powerful radiation sources, so they do not take full advantage of quantum linearity and parallelism. This prompts the formulation of a new approach.

\begin{figure}[h]
    \centering

    \begin{tikzpicture}[scale=3.0]
        \draw[step=1.0,black,thick] (0.0,0.0) grid (1.0,1.0);
        \draw [gray,thick,dashed] [<->] (0.01,0.01) -- (0.99,0.99);
        \draw [gray,thick,dashed] [<->] (0.01,0.99) -- (0.99,0.01);
        \draw [gray,thick,dashed] [<->] (0.5,0.00) -- (0.5,1.00);
        \draw [gray,thick,dashed] [<-] (0.00,0.5) -- (0.5,0.5);
        \draw [black,ultra thick] [->] (0.5,0.5) -- (1.00,0.5);
        \draw[step=1.0,black,thick] (1.0,0.0) grid (2.0,1.0);
        \draw [gray,thick,dashed] [<-] (1.01,0.01) -- (1.5,0.5);
        \draw [gray,thick,dashed] [<-] (1.01,0.99) -- (1.5,0.5);
        \draw [gray,thick,dashed] [->] (1.5,0.5) -- (1.99,0.99);
        \draw [gray,thick,dashed] [->] (1.5,0.5) -- (1.99,0.01);
        \draw [gray,thick,dashed] [<->] (1.5,0.00) -- (1.5,1.00);
        \draw [gray,thick,dashed] [<-] (1.00,0.5) -- (1.5,0.5);
        \draw [gray,thick,dashed] [->] (1.50,0.5) -- (2.00,0.5);
    \end{tikzpicture}

    \begin{tikzpicture}[scale=1.0]
        \draw [black!80,thick] [->] (1.5,1.0) -- (1.5,0.0) node[midway,right,overlay] {$\hat U_{P}$};
    \end{tikzpicture}
    
    \begin{tikzpicture}[scale=3.0]
        \draw[step=1.0,black,thick] (0.0,0.0) grid (1.0,1.0);
        \draw [gray,thick,dashed] [<->] (0.01,0.01) -- (0.99,0.99);
        \draw [gray,thick,dashed] [<->] (0.01,0.99) -- (0.99,0.01);
        \draw [gray,thick,dashed] [<->] (0.5,0.00) -- (0.5,1.00);
        \draw [gray,thick,dashed] [<-] (0.00,0.5) -- (0.5,0.5);
        \draw [gray,thick,dashed] [->] (0.5,0.5) -- (1.00,0.5);
        \draw[step=1.0,black,thick] (1.0,0.0) grid (2.0,1.0);
        \draw [gray,thick,dashed] [<-] (1.01,0.01) -- (1.5,0.5);
        \draw [gray,thick,dashed] [<-] (1.01,0.99) -- (1.5,0.5);
        \draw [gray,thick,dashed] [->] (1.5,0.5) -- (1.99,0.99);
        \draw [gray,thick,dashed] [->] (1.5,0.5) -- (1.99,0.01);
        \draw [gray,thick,dashed] [<->] (1.5,0.00) -- (1.5,1.00);
        \draw [gray,thick,dashed] [<-] (1.00,0.5) -- (1.5,0.5);
        \draw [black!80,ultra thick] [->] (1.50,0.5) -- (2.00,0.5);
    \end{tikzpicture}

    \begin{tikzpicture}[scale=1.0]
        \draw [black!80,thick] [->] (1.5,1.0) -- (1.5,0.0) node [midway,right,overlay] {$\hat U_{AR}$};
    \end{tikzpicture}
    
    \begin{tikzpicture}[scale=3.0]
        \draw[step=1.0,black,thick] (0.0,0.0) grid (1.0,1.0);
        \draw [gray,thick,dashed] [<->] (0.01,0.01) -- (0.99,0.99);
        \draw [gray,thick,dashed] [<->] (0.01,0.99) -- (0.99,0.01);
        \draw [gray,thick,dashed] [<->] (0.5,0.00) -- (0.5,1.00);
        \draw [gray,thick,dashed] [<-] (0.00,0.5) -- (0.5,0.5);
        \draw [gray,thick,dashed] [->] (0.5,0.5) -- (1.00,0.5);
        \draw[step=1.0,black,thick] (1.0,0.0) grid (2.0,1.0);
        \draw [gray,thick,dashed] [<-] (1.01,0.01) -- (1.5,0.5);
        \draw [gray,thick,dashed] [<-] (1.01,0.99) -- (1.5,0.5);
        \draw [black!40,ultra thick] [->] (1.5,0.5) -- (1.99,0.99);
        \draw [black!40,ultra thick] [->] (1.5,0.5) -- (1.99,0.01);
        \draw [gray,thick,dashed] [<->] (1.5,0.00) -- (1.5,1.00);
        \draw [gray,thick,dashed] [<-] (1.00,0.5) -- (1.5,0.5);
        \draw [black!60,ultra thick] [->] (1.50,0.5) -- (2.00,0.5);
    \end{tikzpicture}

    \caption{A visualization of the angular redistribution method based on radiation travelling along a single direction. Each arrow corresponds to an angular direction vector along which radiation can travel. Dashed gray arrows correspond to directions with no radiation intensity at that step of the simulation. Radiation travelling entirely along the horizontal axis in the left grid cell (bolded black arrow) is first propagated to the next grid cell by $\hat U_P$ and then redistributed across the adjacent grid cells (less bolded gray arrows) by $\hat U_{AR}$.}
    
    \label{fig:angular_redistribution_visual}
\end{figure}
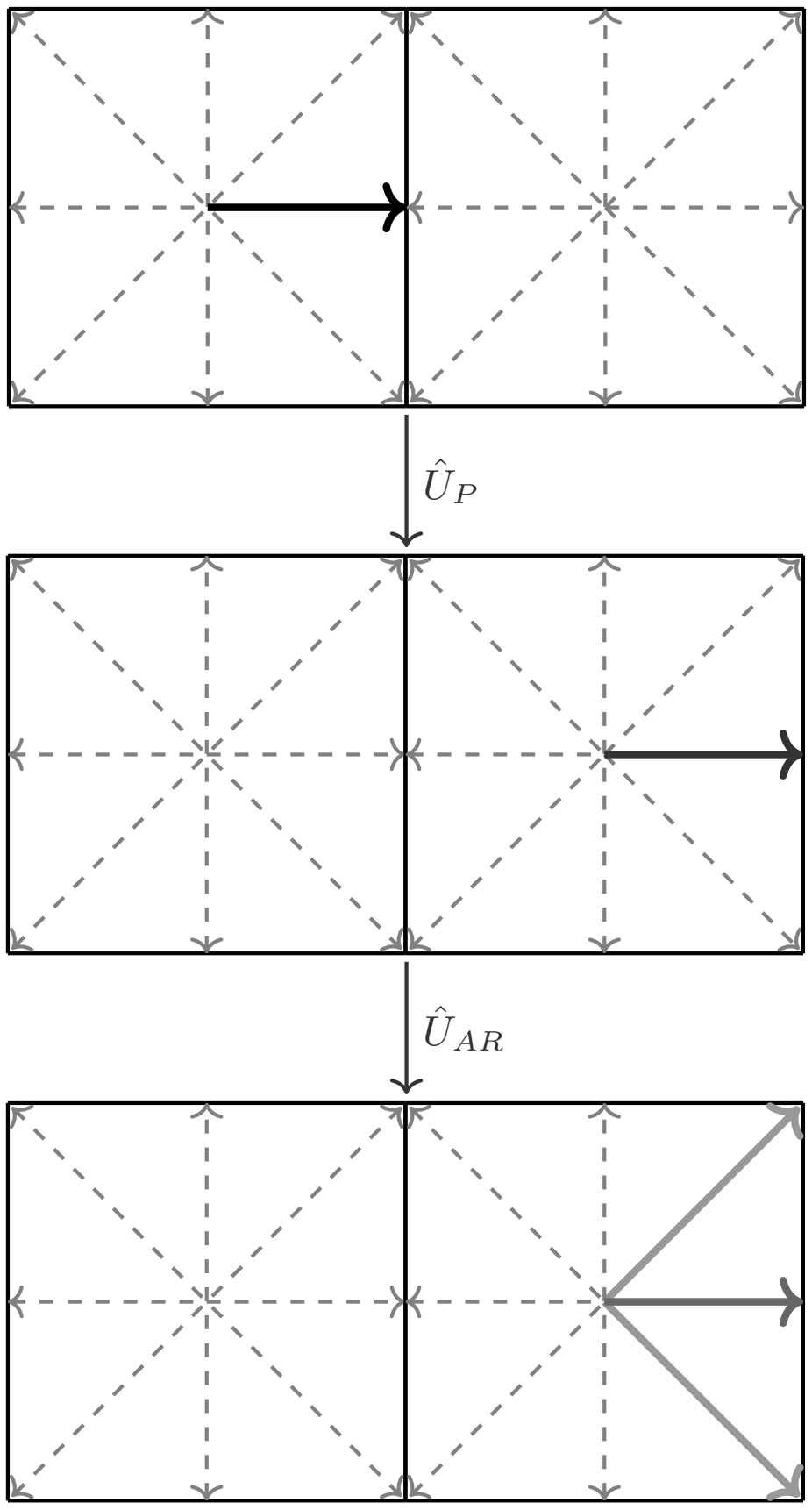

To resolve this problem, our approach, which we term angular redistribution, fills the gaps by propagating a desired fraction of the radiation along adjacent directions, where physically one would expect some amount to propagate were there a direction vector along that angle. The essence of this method is visualized in Figure \ref{fig:angular_redistribution_visual}.

Most importantly, radiation is not only propagated along its typical direction as according to the traditional LBM formulation, but rather it is split between the central direction and adjacent directions. In reality, we expect that the radiation near the boundaries of the arc around the central direction will actually propagate spatially similarly to radiation along the next discrete angular direction. In the near-field, this angular displacement of the simulated radiation from its true position is minimal, but it grows linearly with the distance from the source, making this discrepancy significant.

A similar idea was used by the authors of Ref.~\cite{Coelho2002}, who instead apply a \textit{spatial} redistribution; our angular redistribution method is more robust because it only fills in adjacent rays along the same approximate direction at the same lattice site, thus better preserving flux, and it is significantly cheaper as a quantum algorithm since it does not need to act on the spatial lattice qubits, which are far greater in number than the angular direction qubits.

The angular redistribution methodology results in the following modified form of the LBM radiative transfer equation:
\begin{eqnarray*}
    I_\mu(x \pm \delta_x, t + \delta_t) - I_\mu(x, t) = \sum_{\nu=0}^{m-1} \alpha_{\mu\nu} I_\nu(x, t),
\end{eqnarray*}
where we fix a primary direction $\mu$ and add to it a fraction $\alpha_{\mu\nu}$ of the radiation intensity at that site travelling in the $\nu$ direction, given by $I_\nu$.

There are multiple possible approaches for the formulation of angular redistribution as a quantum operator. Under one approach, one applies a quantum walk unitary to the direction register, which is effectively nearest-neighbors scattering; this is the approach we focus on in this work.

In order to broaden the generality of this approach, we use a mask qubit to indicate which lattice sites should experience angular redistribution; this allows us to both accommodate sources other than stars and also constrain the side-effects of angular redistribution far from sources. This approach has the advantage of being relatively efficient since it only requires operations between switch, direction, mask, and ancilla qubits and most notably not the lattice qubits.

For a nearest-neighbor angular redistribution scheme which is constant at all lattice sites, the operator takes the following representation:
\begin{align*}
    \hat U_{AR} &= \hat V_\alpha^\dagger \hat V_{AR} \hat V_\alpha,
\end{align*}
where $\hat V_\alpha$ prepares a special superposition of the two ancilla qubits used for angular redistribution, and $\hat V_{AR}$ performs the angular redistribution using operators controlled on the two ancillas.

We discuss $\hat V_{AR}$ first. This operator takes the following representation:
\begin{widetext}
    \begin{align*}
        \hat V_{AR}
            &= \I_{\overline{a} l} \otimes \ket{1}_m\bra{1}_m \otimes \ket{0}_s\bra{0}_s \otimes \bigg[\bigg(\ket{00}_{a}\bra{00}_{a} + \ket{11}_{a}\bra{11}_{a}\bigg) \otimes \hat{R}_0 + \ket{01}_{a}\bra{01}_{a} \otimes \hat{R}_+ + \ket{10}_{a}\bra{10}_{a} \otimes \hat{R}_-\bigg] \\
            &\ + \I_{\overline{a} l} \otimes \ket{0}_m\bra{0}_m \otimes \ket{1}_s\bra{1}_s \otimes \I_{a} \otimes \I_d,
    \end{align*}
\end{widetext}
where $a$ indicates the ancilla qubits chosen for angular redistribution and $\overline{a}$ indicates the remaining ancilla qubits which are irrelevant to this step. In our case, we choose these to be the final two qubits in the quantum circuit diagrams, when read top-to-bottom.

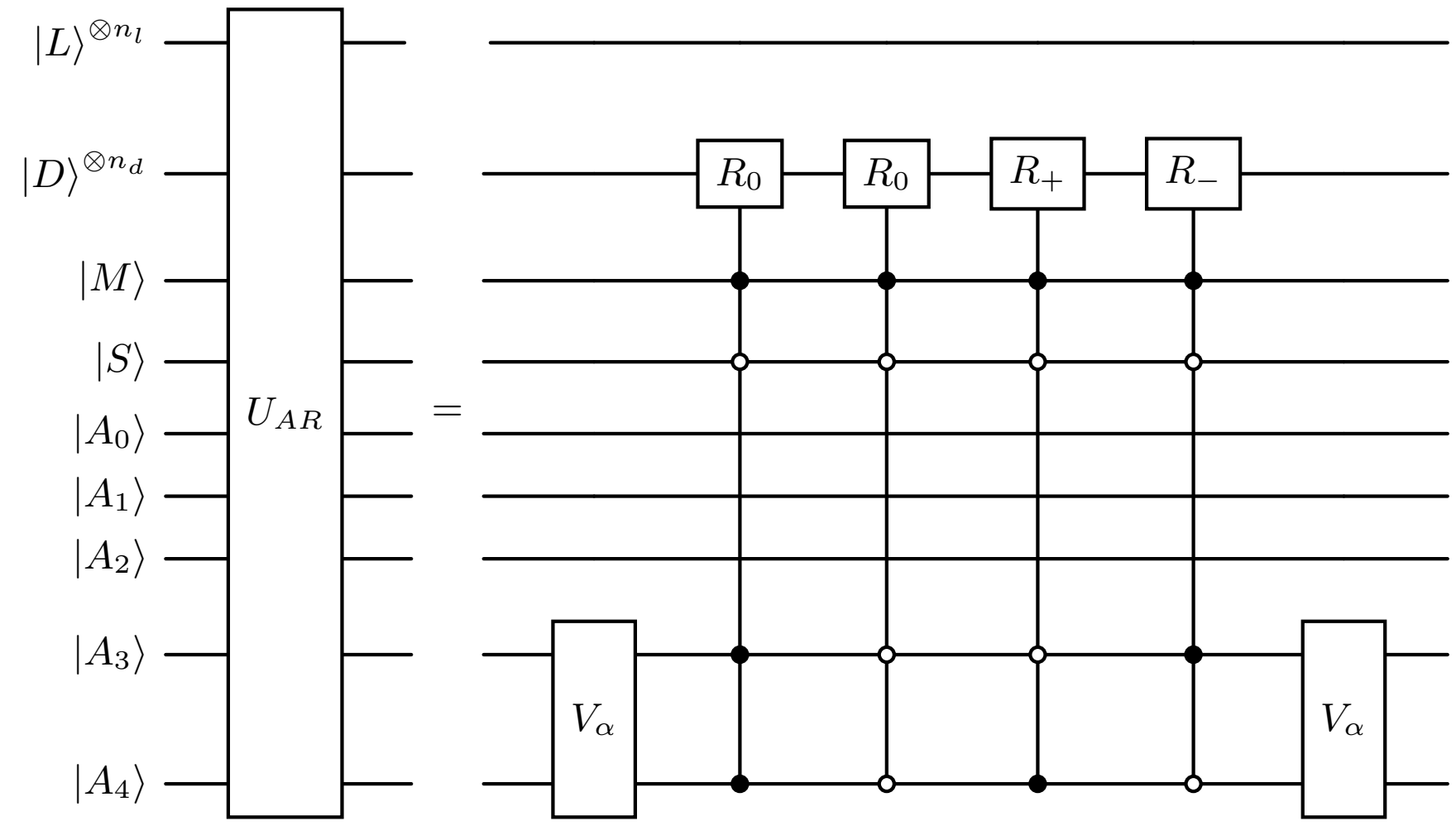
\begin{figure*}[ht!]
    \begin{quantikz}
        \lstick{$\ket{L}^{\otimes n_l}$}
            & \gate[9]{U_{AR}}
            & \midstick[9,brackets=none]{=}
            &
            &
            &
            & 
            &
            &
            &
        \\
        \lstick{$\ket{D}^{\otimes n_d}$}
            &
            &
            &
            & \gate[1]{R_0}
            & \gate[1]{R_0}
            & \gate[1]{R_+}
            & \gate[1]{R_-}
            & 
            &
        \\
        \lstick{$\ket{M}$}
            &
            &
            & 
            & \ctrl{0} \wire[u]{q}
            & \ctrl{0} \wire[u]{q}
            & \ctrl{0} \wire[u]{q}
            & \ctrl{0} \wire[u]{q}
            & 
            & 
        \\
        \lstick{$\ket{S}$}
            &
            &
            & 
            & \octrl{0} \wire[u]{q}
            & \octrl{0} \wire[u]{q}
            & \octrl{0} \wire[u]{q}
            & \octrl{0} \wire[u]{q}
            & 
            & 
        \\
        \lstick{$\ket{A_0}$}
            &
            &
            &
            & \wire[u]{q}
            & \wire[u]{q}
            & \wire[u]{q}
            & \wire[u]{q}
            & 
            &
        \\
        \lstick{$\ket{A_1}$}
            &
            &
            &
            & \wire[u]{q}
            & \wire[u]{q}
            & \wire[u]{q}
            & \wire[u]{q}
            & 
            &
        \\
        \lstick{$\ket{A_2}$}
            &
            &
            &
            & \wire[u]{q}
            & \wire[u]{q}
            & \wire[u]{q}
            & \wire[u]{q}
            & 
            &
        \\
        \lstick{$\ket{A_3}$}
            &
            &
            & \gate[2]{V_{\alpha}}
            & \ctrl{0} \wire[u]{q}
            & \octrl{0} \wire[u]{q}
            & \octrl{0} \wire[u]{q}
            & \ctrl{0} \wire[u]{q}
            & \gate[2]{V_{\alpha}}
            & 
        \\
        \lstick{$\ket{A_4}$}
            &
            &
            & 
            & \ctrl{0} \wire[u]{q}
            & \octrl{0} \wire[u]{q}
            & \ctrl{0} \wire[u]{q}
            & \octrl{0} \wire[u]{q}
            & 
            & 
        \\
    \end{quantikz}
    \caption{The quantum circuit for the angular redistribution step for a $D2Qm$ scheme. Shown is nearest-neighbor coupling; $\hat V_\alpha$ performs a Householder project onto the angular redistribution coefficients vector and the controlled gates apply the nearest-neighbor redistribution along the $\mu$ and $\mu\pm1$ directions; the $R_0$ gate is equivalent to the identity in all cases and thus can be skipped entirely, but we visualize it here for completeness. This circuit can be adapted for more complex angular redistribution schemes by modifying the ancilla state preparation unitary $\hat V_\alpha$ and the redistribution operators $\hat R_i$.}
    \label{fig:angular_redistribution_circuit}
\end{figure*}

The first set of terms in the square braces selects intensity terms via the projector on the switch qubit and redistributes terms probabilistically: the first term in square braces applies $\hat R_0$ to terms whose angular redistribution ancillas are either in the $00$ or $11$ states; the second term in square braces applies $\hat R_+$ to terms whose angular redistribution ancillas are in the $01$ state; the third term in square braces applies $\hat R_-$ to terms whose angular redistribution ancillas are in the $10$ state. The first redistribution operator is simply the identity and corresponds to the action of keeping some of the radiation intensity travelling along its original direction. The non-trivial redistribution operators $\hat R_\pm$ are defined as follows:
\begin{align*}
    \hat R_\pm &= \sum_{\nu=0}^{m-1} \ket{\mu \pm 1 \mod{m}}_d \bra{\mu}_d.
\end{align*}
The circuit for $\hat U_{AR}$ is shown in Figure \ref{fig:angular_redistribution_circuit}.

It remains to discuss the form of the $\hat V_\alpha$ operator. In the case of angular redistribution which is uniform over initial propagation directions, we construct this gate using a Householder transform: let $\vec a$ be our desired angular redistribution coefficient vector, such that $a_i = \alpha_{\mu,\mu+i}$ is a single column of the angular redistribution tensor, and define $q_i = \delta_{i,0} - a_i$; then, our state preparation gate is the Householder projection operator
\begin{align*}
    \hat V_\alpha &= 1 - 2 \dfrac{q_i q_i^T}{\norm{q_i}^2}.
\end{align*}

Finally, determination of the appropriate coefficients $\alpha_{\mu\nu}$ is largely dependent upon the geometry of the grid, the angular divisions of the lattice scheme, and the desired behavior in local regions. We have found that, for many simulations involving isotropic sources such as stars, the most efficient angular redistribution scheme is as follows:
\begin{align*}
\alpha_{\mu\nu}
    &= \begin{cases}
        \frac{2}{m} & \nu=\mu \\
        \left(1 - \frac{2}{m}\right)/2 & \nu=\mu\pm1 \\
        0 & \mathrm{otherwise}
    \end{cases}
\end{align*}

Although there may be better redistribution parameters, we have found both geometrically and empirically that the above values minimize the anisotropy. This is shown in Figure \ref{fig:angular_redistribution_circuit}.

If all sources have no propagation oblique to the lattice directions (such as radial propagation), then angular redistribution is technically unnecessary in non-scattering media. However, certain systems such as rectangular plates may have fringe effects at corners, which are not captured by the lattice directions and thus necessitate angular redistribution.

If some sources have curvature, but others do not, then it is possible to only apply angular redistribution in the vicinity of the curved surfaces. In this case, one defines the spatially-varying angular redistribution \textit{functions} $\alpha_{\mu\nu}(\vec x)$.
This would increase computational complexity, however, as the angular redistribution step would have to be controlled separately for each set of lattice sites with the same angular redistribution values.

This approach is particularly advantageous over interpolation because a quantum interpolation operator would likely require many gates involving lattice and direction qubits along the interpolation stencil, whereas our approach can be applied globally without involving any lattice qubits.
Even if one must implement local angular redistribution at select lattice sites, our approach only requires coupling between two lattice qubits and up to two data qubits.
Furthermore, there cannot be any guarantee that interpolation conserves quantities such as photon number, because interpolation formulae are approximate and the radiation quantities which are filled in are not taken from existing radiation. One is instead assuming that a certain quantity of radiation should exist in a location and inserting it without explicitly tracking how much radiation should exist at that site in that direction. Although our approach is also inexact in how radiation is redistributed, we can guarantee conservation because all redistributed radiation is subtracted from neighboring regions of the simulation, so we neither add or remove radiation in-place.
Moreover, local interpolation would only fill gaps in empty lattice sites which are in close proximity to non-empty lattice sites, which would not be as effective far from sources where the gaps span a large region in space.

It may be possible to unify this step with a scattering step, but existing formulations of scattering, such as that detailed in Ref.~\cite{Igarashi2024}, assume isotropic scattering from each individual direction to every other direction \cite{Igarashi2024}, which is not sufficient for our angular redistribution approach. In fact, our angular redistribution circuit implements nearest-neighbor scattering and could be extended to obtain a more general scattering circuit. However, such an approach would have the same obstacle of preparing orthogonal states with coefficients $\sigma_{\mu\nu}$.

\subsection{\label{sec:measurement} Measurement}

At the end of each timestep, one must measure the full state of the quantum system and perform post-processing of the results, generally as follows:
\begin{enumerate}
    \item From the measurement counts for each state, one must post-select the states for which the ancilla qubits are all in the $\ket{0}$ state. In practice for a simulation with all steps except for global absorption and scattering, we find that the total number of post-selected counts tends to be around 25\% of the total counts. This is the typical setting for cosmological simulation, which usually feature spatially-varying absorption and scattering.
    \item One may also post-select further to apply special boundary conditions, such as absorbing boundary conditions. In the special case where the absorption is either exactly 0 or 1 at each lattice site, one may simply ignore the results of measurements on such lattice sites.
    \item Once the post-selected counts have been obtained, one computes the frequencies with respect to the original total counts before post-selection.
    \item Then, one obtains the state amplitudes, which are the intensities and sources, by taking the square roots of the frequencies.
    \item Finally, the system is prepared in the post-selected state, and another timestep is performed.
\end{enumerate}
The greatest potential for both accuracy and runtime improvements lies at this stage of the algorithm. We attempted to reduce the number of measurements and redundancy in the simulations by conditioning each successive timestep of the simulation on measurements of the previous ancilla qubits; this approach would allow for multiple, and even all, timesteps to be run continuously without breaking for measurements, post-selection, and state re-preparation.
Although it would not provide access to the full system state between timesteps within a continuous run, this can be leveraged by performing measurements only when this intermediate output is required, such as for input to chemistry or hydrodynamics solvers. We have found that classically simulating the repeated mid-circuit measurements and conditional operations makes this approach much slower than the explicit post-selection cycle; however, it remains to be tested on real quantum hardware, where the primary concerns would be the speed and fidelity of such measurements and conditional operations.

\section{\label{sec:impl} Implementation}

We implement the algorithm using the Python language and the Qiskit and Qiskit Aer packages \cite{Qiskit}. We found empirically that computation of the propagation and angular redistribution unitary operators tends to be the slowest classical preparation step. We were able to accelerate these steps by computing complex circuits once and serializing them using the \texttt{QPY} protocol implemented in Qiskit. The serialized data can be stored as cache and deserialized later for reuse. The package used to generate the data in this work and its implementation details are discussed further in the companion work \cite{QuaRT-JOSS}; its source code is also publicly available \cite{QuaRT}.

\section{\label{sec:results} Simulation results}

We make use of three fundamental two-dimensional test cases to demonstrate the correctness of this method:
\begin{enumerate}
    \item \textbf{Isotropic source} - demonstrates the anisotropy reduction due to our angular redistribution method relative to traditional LBM methodology
    \item \textbf{Shadow test} - demonstrates the ability of the method to capture shadows of opaque objects
    \item \textbf{Crossing radiation beams} - demonstrates the ability of the method to correctly propagate radiation at an angle oblique to the simulation grid
\end{enumerate}

The scale and resolution of simulations performed in this work is limited most by memory constraints on quantum circuit simulations; we work around this issue by showing multiple simulations at increasingly-large scales to demonstrate convergence. All circuits were simulated exactly to obtain exact probability amplitudes; more practically, we find that shot counts on the order of $10^5$ may be sufficient depending on the problem size desired degree of accuracy \cite{Igarashi2024}.

\subsection{Isotropic source}

We test the degree of isotropy of our method by simulating the radiation from a single point source, which should be spherically-symmetric, both with and without angular redistribution.

\begin{figure}[ht!]
   \centering
   \includegraphics[width=0.45\textwidth]{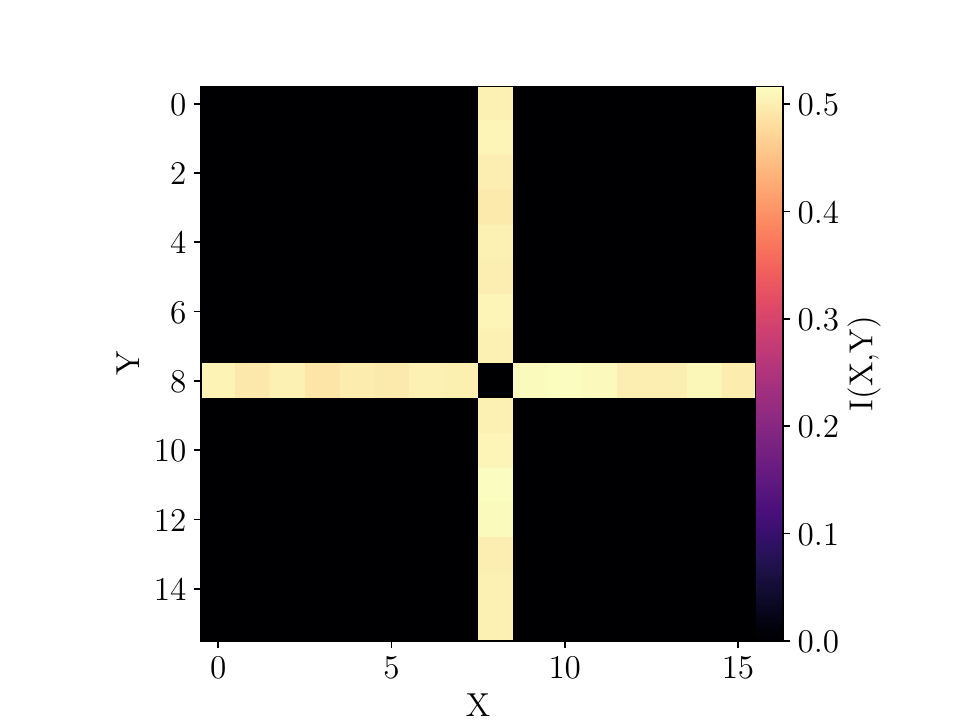}
   \caption{Result of a D2Q4, 16x16 LBM simulation without angular redistribution, showing a lack of angular symmetry.}
   \label{fig:aniso_src_visual}
\end{figure}

To begin, we ran simulations with the traditional lattice Boltzmann methodology with various lattice schemes, which most importantly excludes angular redistribution, one of which is shown in Figure \ref{fig:aniso_src_visual} for D2Q4 on a 16x16 lattice. For every lattice scheme, it is evident that there are large anisotropies which are only improved in part by increasing the number of angular directions.

\begin{figure}[ht!]
   \centering
   \subfigure[D2Q4, 8x8]{\includegraphics[width=0.45\textwidth]{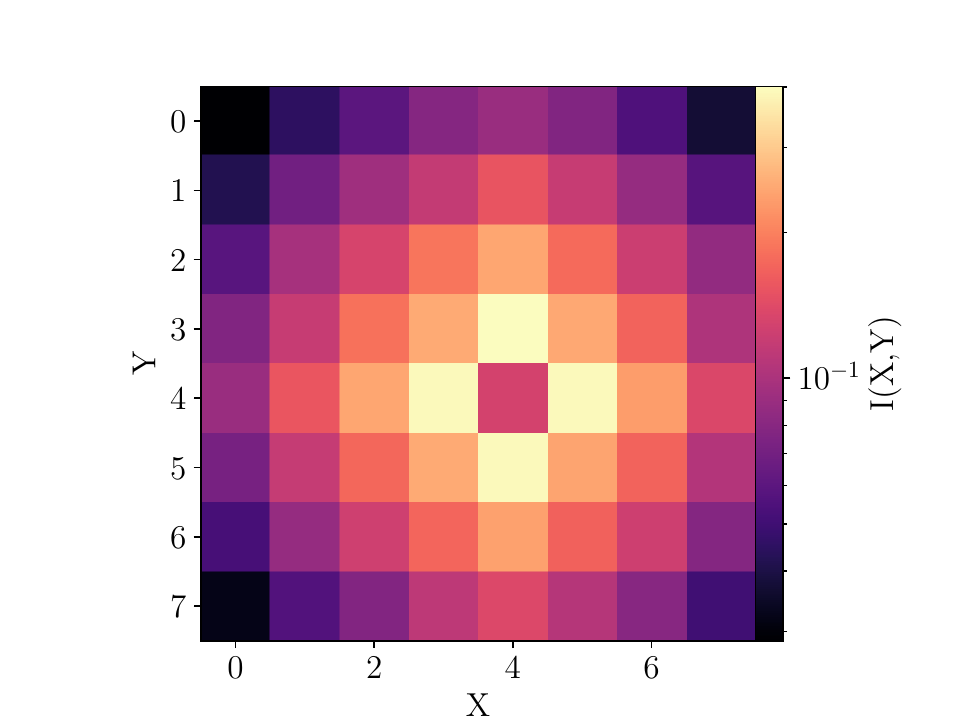}}
   \subfigure[D2Q8, 16x16]{\includegraphics[width=0.45\textwidth]{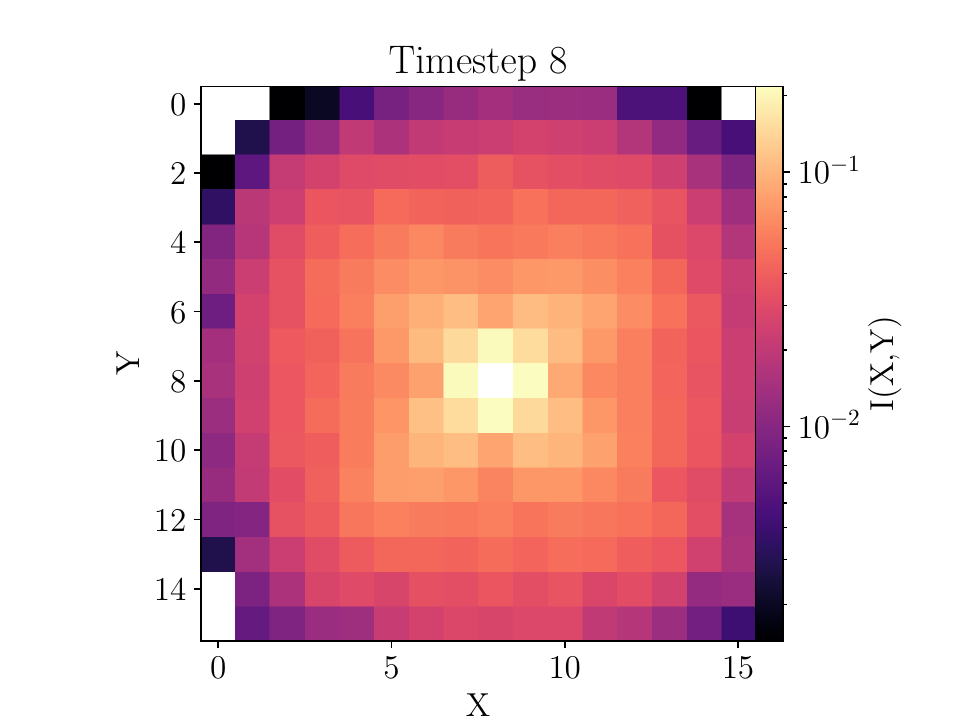}}
   \subfigure[D3Q64, 32x32x32]{\includegraphics[width=0.45\textwidth]{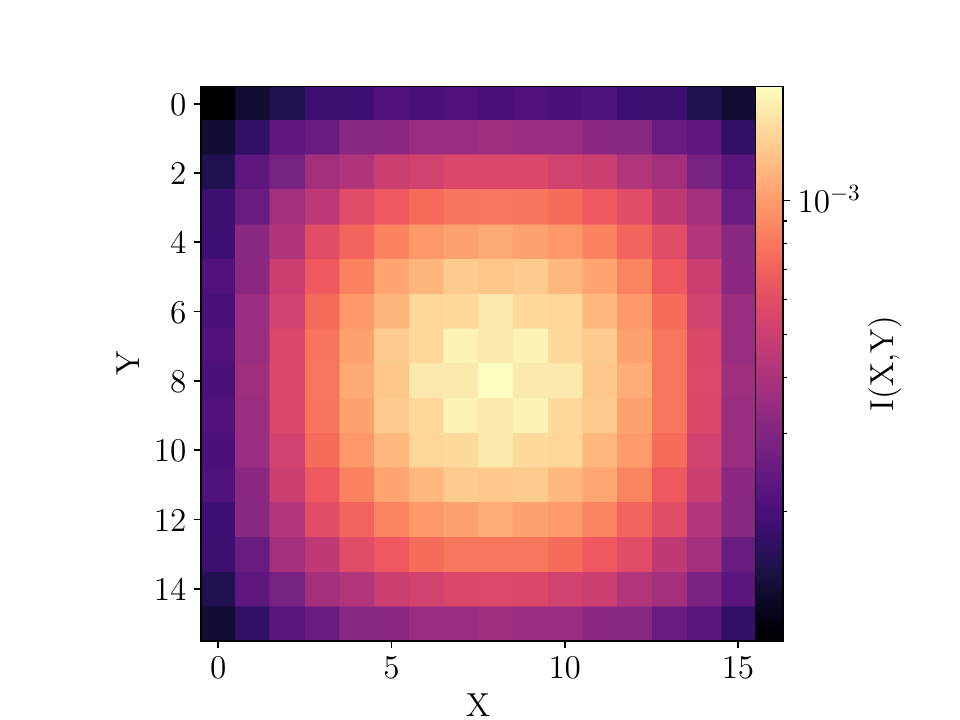}}
   \caption{Results of LBM simulations with angular redistribution for various lattice schemes, showing angular symmetry in each case. Increasing the simulation spatial and angular resolution visually increases the isotropy.}
   \label{fig:iso_src_visual}
\end{figure}

A visual comparison of the simulation results for different LBM schemes with angular redistribution is shown in Figure \ref{fig:iso_src_visual}. The three plots correspond to D2Q4 8x8, D2Q8 16x16, and D3Q64 32x32x32 simulations, respectively, and show the increasing isotropy as the simulation parameters are scaled. In each case, the source has radiation intensity $I=1$ (summed over all directions) which is then radiated around the simulation. We use absorbing boundary conditions so that radiation exiting the simulation box does not re-enter and distort the isotropy study. For the three-dimensional simulation, we only plot the $z=L_z/2$ slice.

\begin{figure*}[ht!]
    \centering
    \subfigure[D2Q4 8x8]{\includegraphics[width=0.4\textwidth]{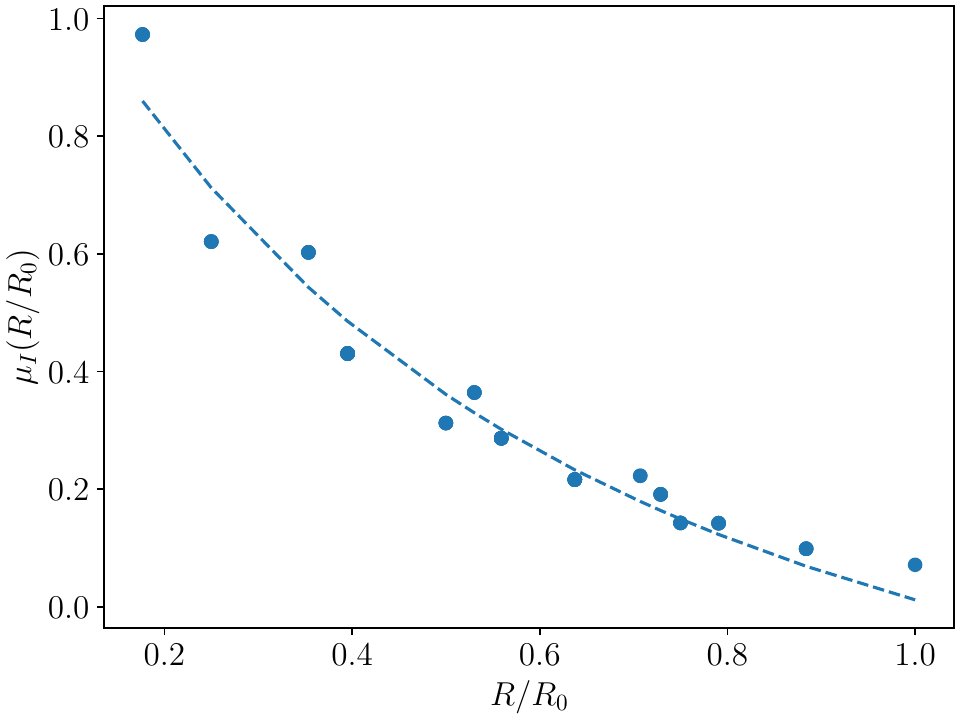}}
    \subfigure[D2Q8 16x16]{\includegraphics[width=0.4\textwidth]{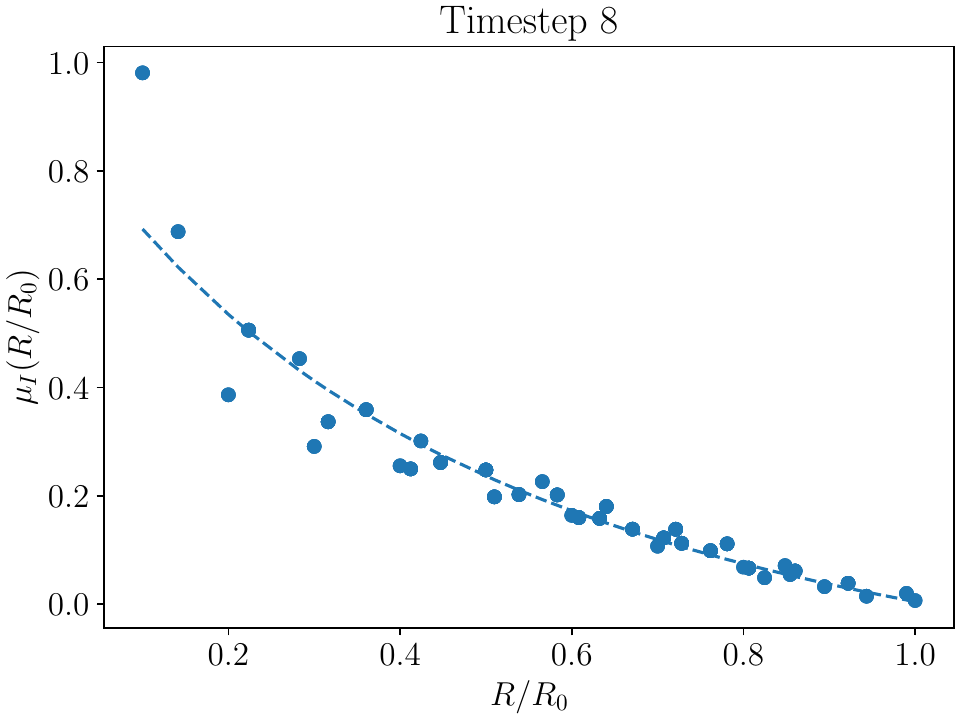}}
    \subfigure[D3Q64 32x32x32]{\includegraphics[width=0.4\textwidth]{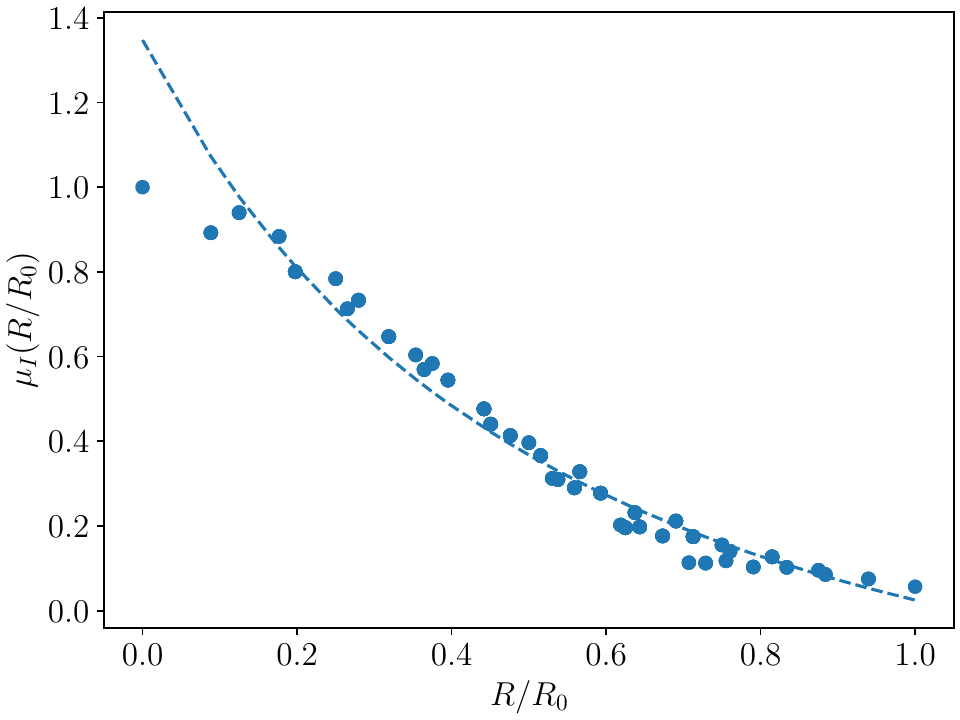}}
    \caption{The mean radial intensity profile for the three lattice schemes, demonstrating the minimal anisotropy with our method. We fit the data to an inverse square law to show that, except for the data right at the origin and at the corners of the grid, the methodology preserves the expected behavior. In the case of subfigure (a), this is where the effect of simulating a spherical source on a Cartesian grid over more than one light-crossing time becomes evident; the corners have a greater distance from the origin than the interiors of the edges, so there is slightly more error at high radii.}
    \label{fig:iso_src_mean_intensity_profile}
\end{figure*}

In Figure \ref{fig:iso_src_mean_intensity_profile}, we plot the mean radial profile of the radiation intensity around the source. The means are computed by collecting intensities which are at the same radius and taking the angular mean at each radius. Due to lattice discretization, there is additional variance in the radial profiles not due to the angular redistribution, but the magnitude of these deviations decrease with system size.

To quantify the degree of anisotropy, we found that the standard deviation tends to be the most reliable measure.

\begin{figure*}[ht!]
    \centering
    \subfigure[D2Q4 8x8]{\includegraphics[width=0.4\textwidth]{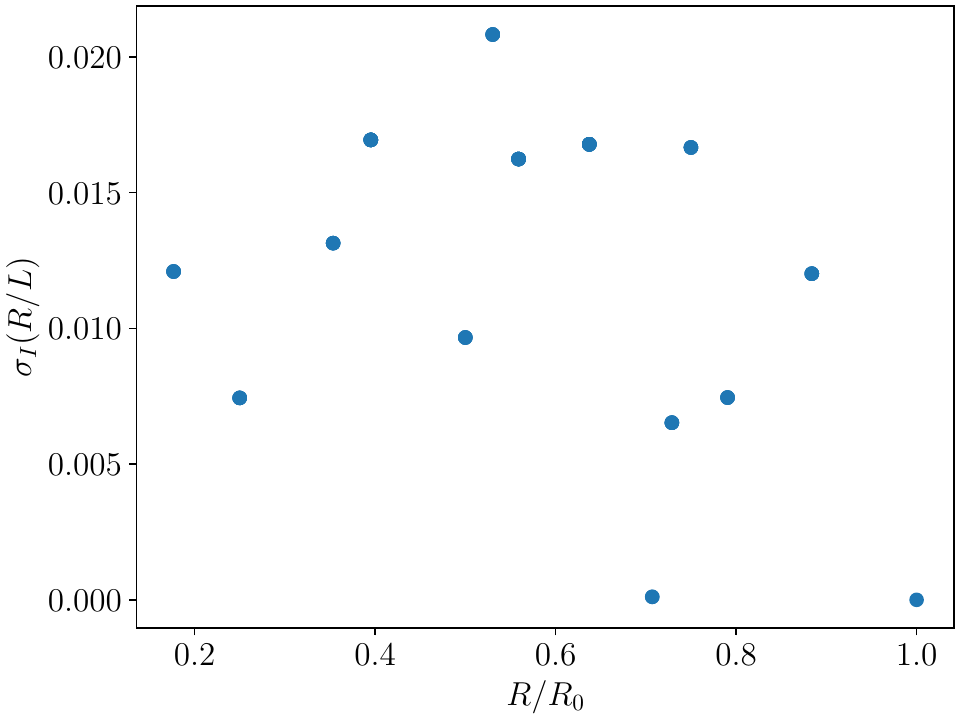}}
    \subfigure[D2Q8 16x16]{\includegraphics[width=0.4\textwidth]{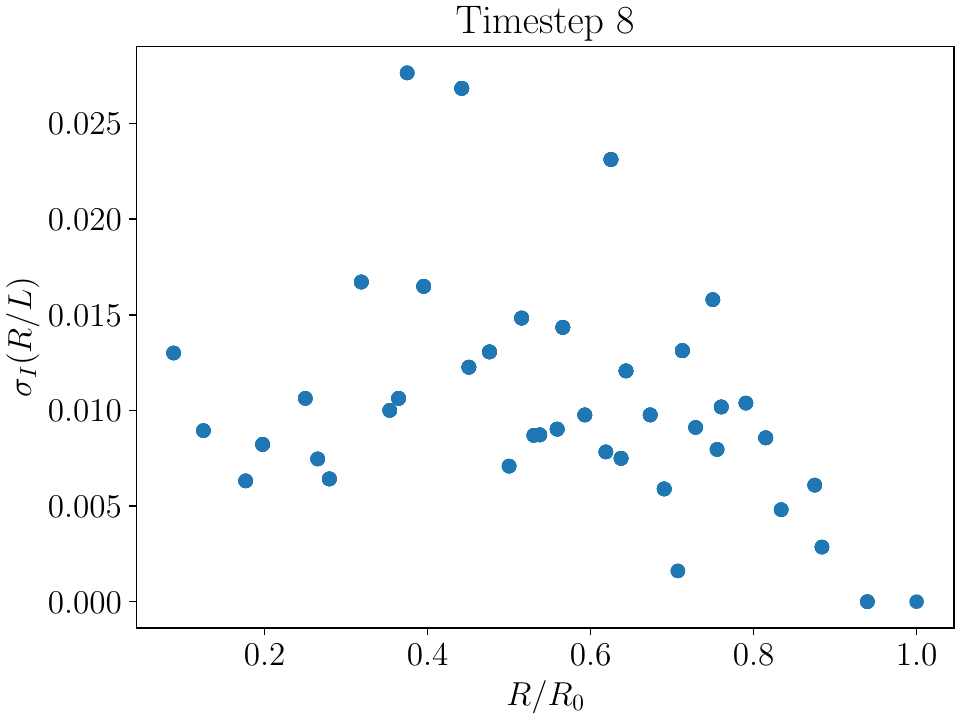}}
    \subfigure[D3Q64 32x32x32]{\includegraphics[width=0.4\textwidth]{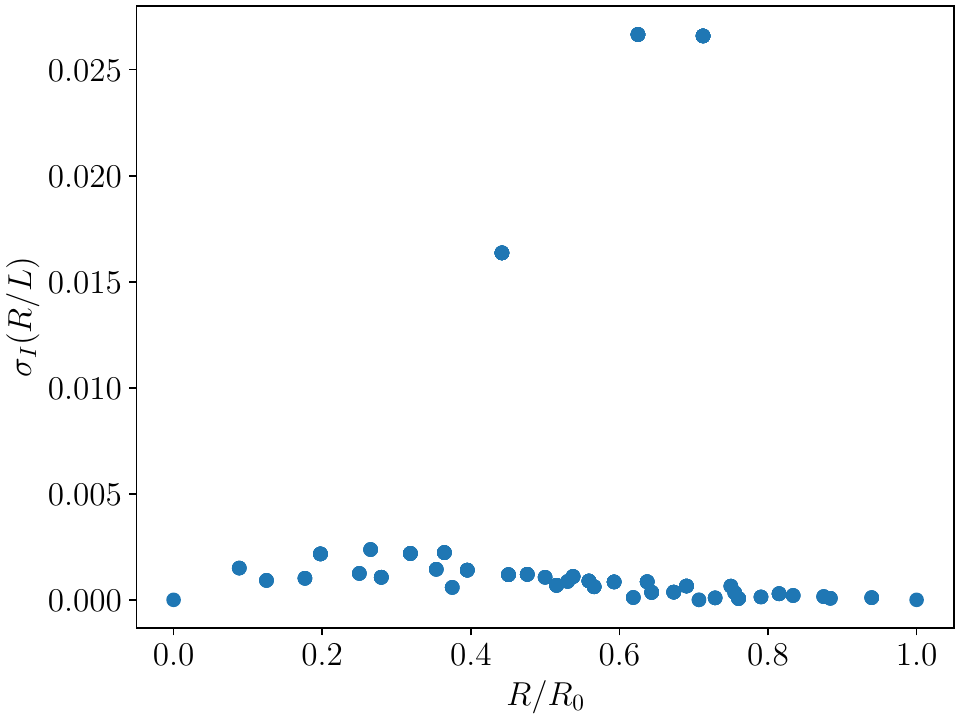}}
    \caption{A radial coefficient of variation profile of the radiation around the source for the three lattice schemes.}
    \label{fig:iso_src_anisotropy_profile}
\end{figure*}

A radial profile of the anisotropy is shown in Figure \ref{fig:iso_src_anisotropy_profile}.
The isotropy of our results is limited largely by the limited resolution accessible through classical simulations of the quantum algorithm. We are able to attain anisotropies below 3\% in larger fully-classical simulations using the same angular redistribution methodology, indicating that the quantum algorithm implementation is capable of achieving even better isotropy on real quantum hardware, where classical computing power is not a limitation.

\subsection{Shadowing}

The shadow test, inspired by the work discussed in Refs. \cite{Iliev2006,Iliev2009,Hayes2003,Jiang2014}, test how well a radiation transport simulation captures shadows | i.e., regions where radiation intensity should be low due the presence of an opaque region which absorbs radiation from a nearby source.

The shadow test is necessary for our angular redistribution methodology because it is possible for radiation to unrealistically wrap around an opaque region through repeated angular redistribution. For example, with a D2Q4 lattice scheme, some of the radiation travelling along the $+x$ direction is redistributed to the $+y$ direction after one timestep, and some of this radiation is further redistributed to the $-x$ direction after a second timestep.

More generally, in a D2Q$m$ scheme with approximately regularly-spaced direction vectors (up to the limitations of the Cartesian grid geometry), it would take approximately $\lceil \frac{m}{2} \rceil$ iterations for radiation travelling along a single direction to completely reverse direction. Assuming constant angular redistribution $\alpha$ at each step and each grid point, this corresponds to a cumulative redistribution factor of $\alpha^{\lceil \frac{m}{2} \rceil}$. For example, a D2Q8 scheme with $\alpha = \frac12$ would have a cumulative redistribution factor of $\frac{1}{16}$, indicating that the radiation intensity behind an opaque cloud should be more than order of magnitude weaker than the radiation intensity in front of the opaque cloud.

We choose our source term to be a single isotropic point source placed to the top of the opaque region. For the opaque region, we set absorbing boundary conditions to all lattice sites $(x, y)$ satisfying $\frac{(x-h)^2}{a^2} + \frac{(y-k)^2}{b^2} \leq 1$, where $h = -N_x/8 = -2$, $k = 0$, $a = N_x/6 = 8/3$, and $b = N_y/3 = 16/3$. We use an isotropic source near the top of the simulation grid with radiation intensity $I = 1$ (summed over all directions). This test utilizes angular redistribution to ensure isotropy, and thus it tests the capturing of shadows under our angular redistribution methodology. We set the angular redistribution mask qubit to be $\ket{1}_m$ only for terms within a radius $r = \frac14 (N_x+N_y)/2 = 4.5$. This radius is chosen because it is large enough to leak past the center of the opaque region, testing the performance of the angular redistribution at such proximity to such regions.

\begin{figure}[ht!]
   \centering
   \includegraphics[width=0.4\textwidth]{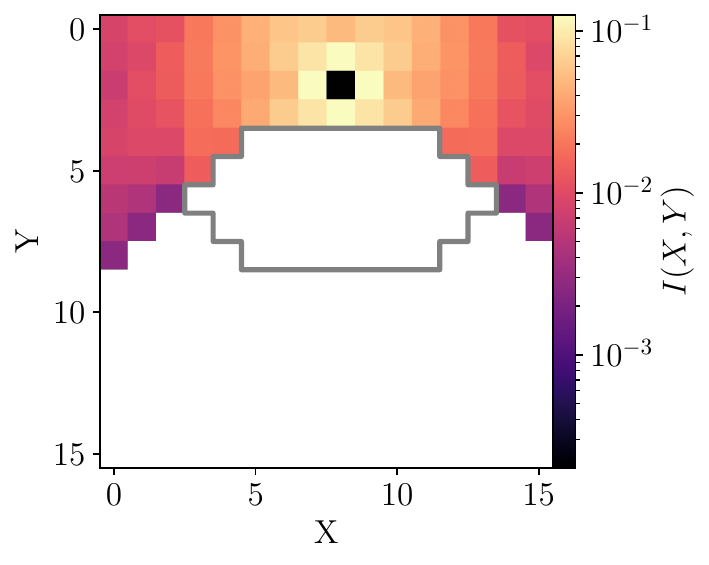}
   \caption{Result of a D2Q8 16x16 shadow point source test with angular redistribution, showing weak shadowing. The colorbar uses a logarithmic norm, so white regions have exactly 0 intensity, allowing us to see by inspection that there is an exact shadow behind the cloud.}
   \label{fig:shadow_point}
\end{figure}

The result of this simulation is shown in Figure \ref{fig:shadow_point}. We see that our method is able to nearly perfectly capture the shadow over eight times as many timesteps as simulation box length, demonstrating the stability of the shadow. The most notable leakage is around the corner of the shadow, where some radiation slightly curls around. However, this plot holds stable for many timesteps, so we did not find this to be a concern for longer simulations. Because the colorbar uses a logarithmic norm, we see from the fact that the entire region behind the cloud is white that there is exactly zero radiation intensity there, confirming that our method passes the shadow test.

\subsection{Crossing radiation beams}

The crossing radiation beams test, inspired by the work discussed in Refs. \cite{Davis2012}, \cite{Jiang2014}, and \cite{Asahina2020}, tests for both ability to stably propagate structured radiation flows at an angle oblique to the simulation grid, as well as the absence of radiation self-interaction over many timesteps,

In the crossing radiation beams test, we set up a 2D simulation grid with dimensions $N_x$ and $N_y$ and periodic boundary conditions along each edge. For the initial conditions, we place two beam sources of radiation at the top of the grid, each of width $w$ at 45 degree angles with respect to the horizontal axis of the simulation domain, facing opposite directions. The left beam (travelling up and to the right) is centered at $L_x/4$, while the right beam (travelling up and to the left) is centered at $3L_x/4$. These parameters are chosen so that, under periodic boundary conditions, radiation exiting through the top of the simulation domain enter at their exact initial locations at the bottom of the simulation domain. In doing so, we are able to simulate the effects of beam crossings over many light-crossing times with minimal cost.

We choose a D2Q8 scheme since this is the simplest scheme which includes direction vectors at 45 degree increments, and any larger scheme would include additional direction vectors which are unnecessary for this test. Since the sources are oblique beams we do not need angular redistribution for this test.

\begin{figure*}
    \centering
    \subfigure{\includegraphics[width=0.9\textwidth]{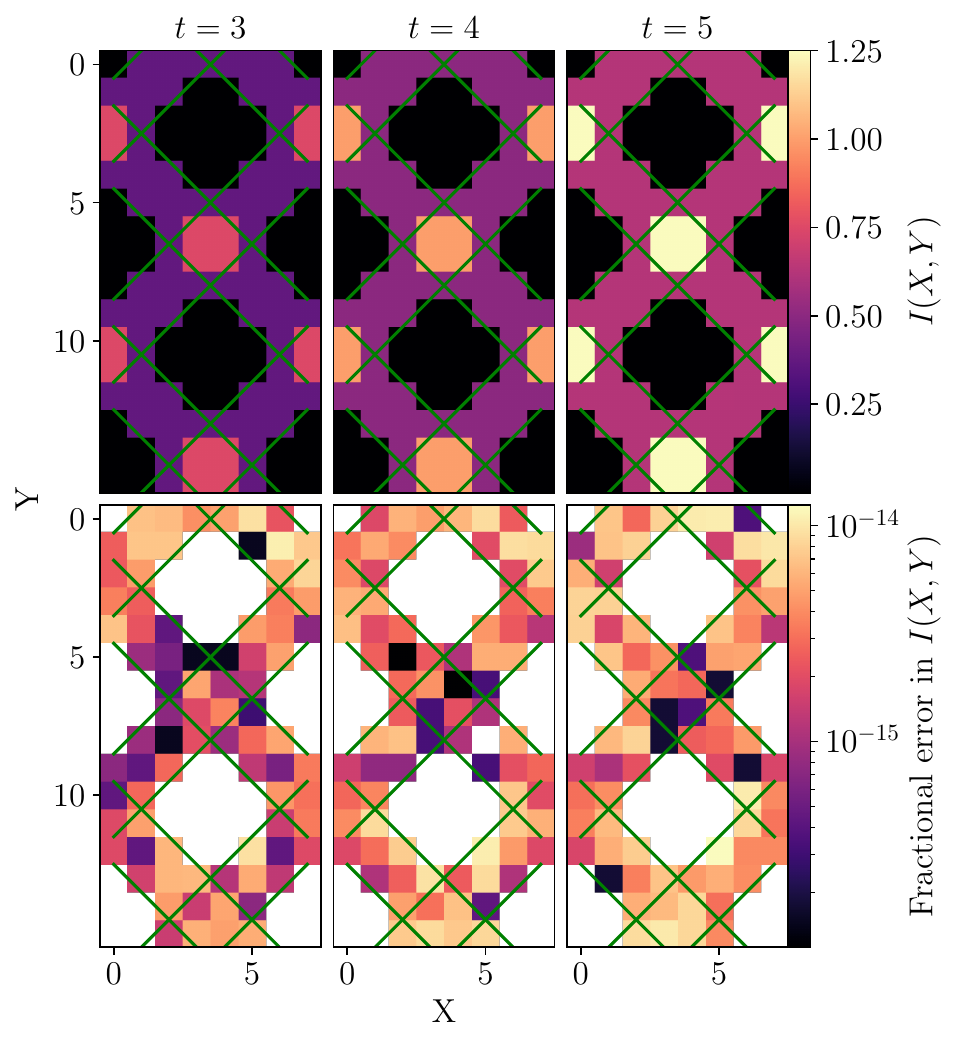}}
    \caption{Result of crossing radiation beam test for a D2Q8 scheme on a 16x8 grid with uniform beams each of total width 3. (Top row) Radiation intensity after $t \in \{3, 4, 5\}$ light crossing times along the longest grid dimension, showing no non-physical collision effects. (Bottom row) Fractional error in radiation intensity relative to exact traces of beam paths through the grid.}
    \label{fig:crb_small_uniform}
\end{figure*}

\begin{figure*}
    \centering
    \subfigure{\includegraphics[width=0.9\textwidth]{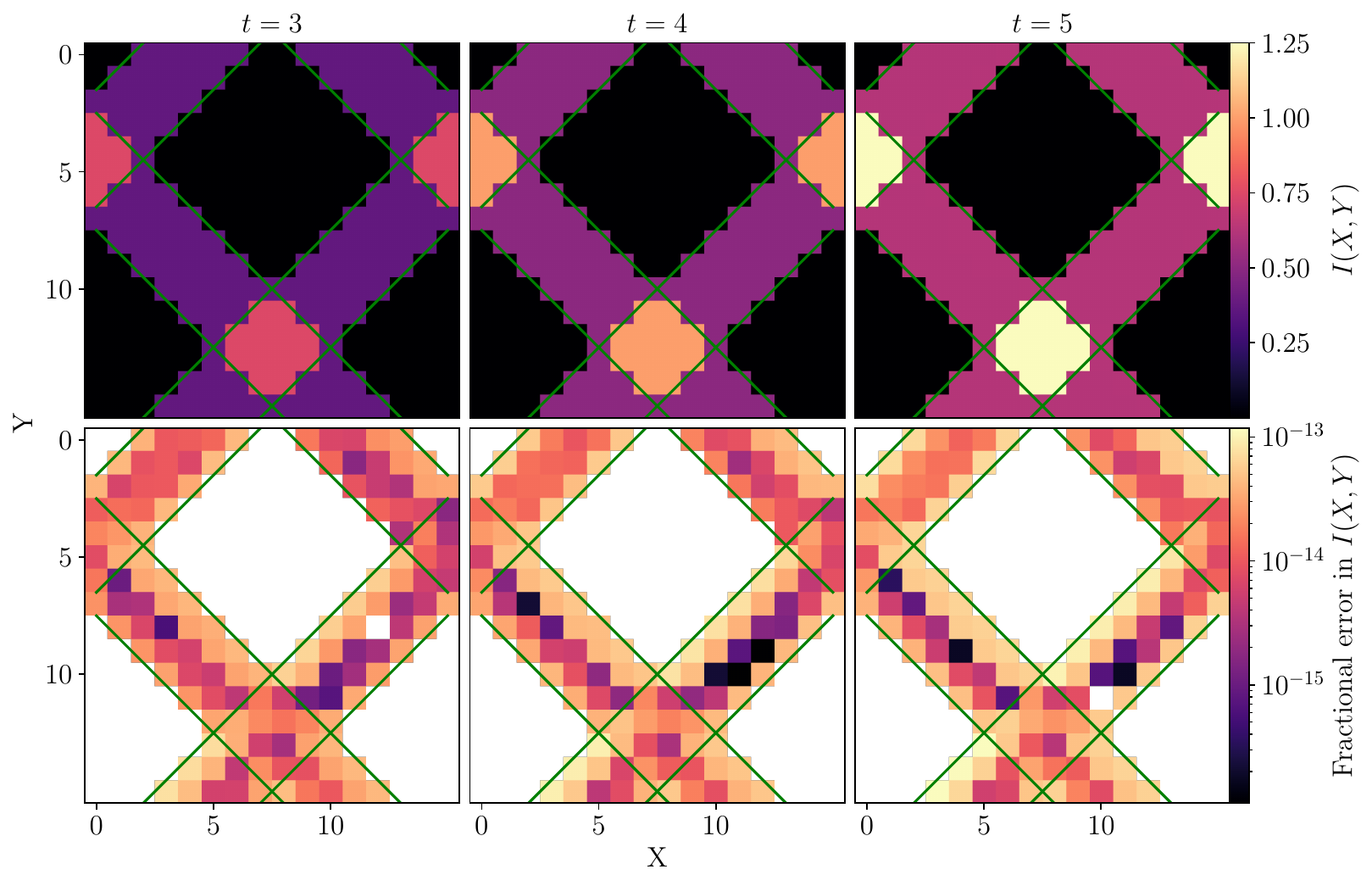}}
    \caption{Result of crossing radiation beam test for a D2Q8 scheme on a 16x16 grid with uniform beams each of total width 5. (Top row) Radiation intensity after $t \in \{3, 4, 5\}$ light crossing times along the longest grid dimension, showing no non-physical collision effects. (Bottom row) Fractional error in radiation intensity relative to exact traces of beam paths through the grid.}
    \label{fig:crb_large_uniform}
\end{figure*}

\begin{figure*}
    \centering
    \subfigure{\includegraphics[width=0.9\textwidth]{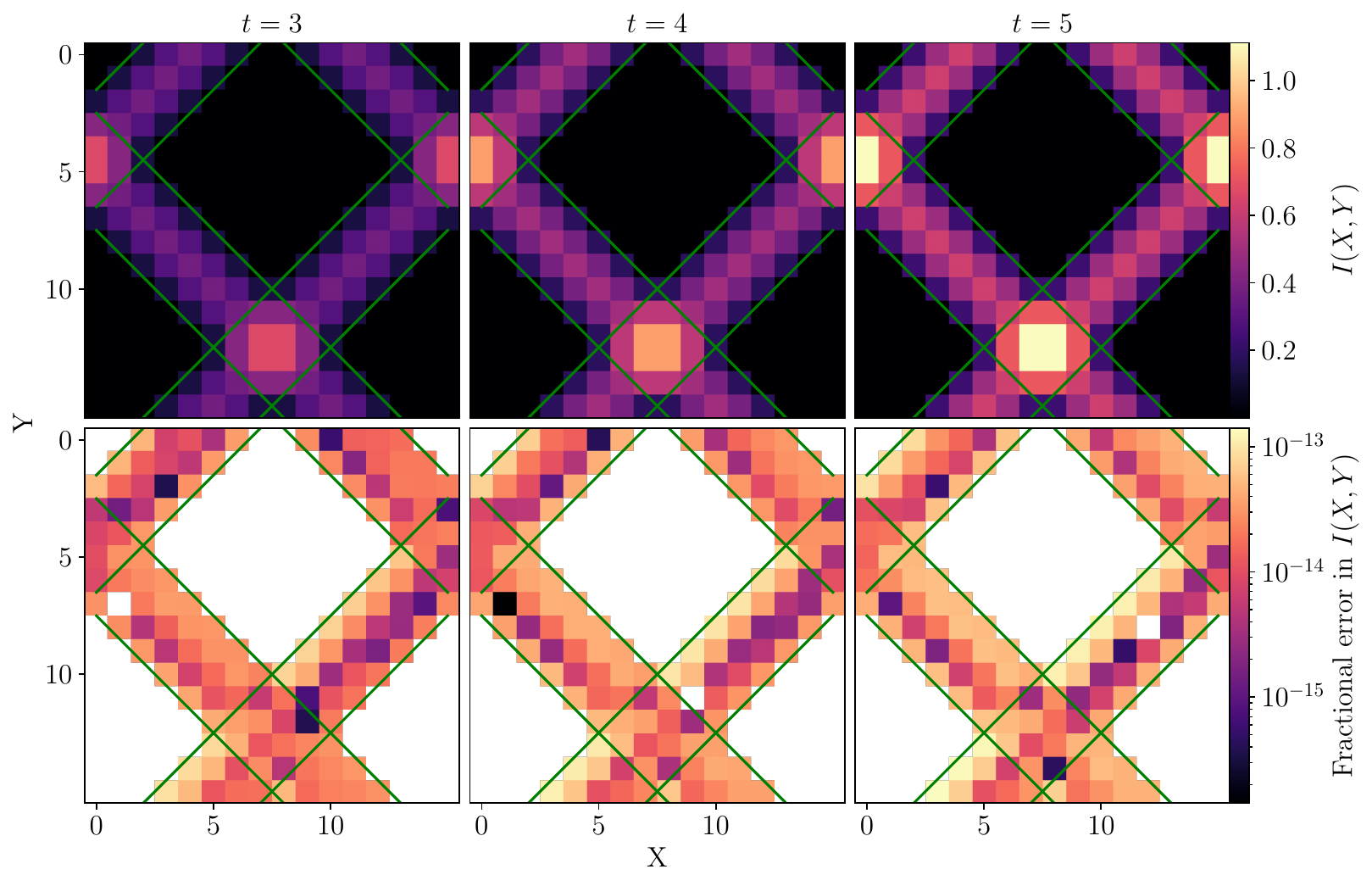}}
    \caption{Result of crossing radiation beam test for a D2Q8 scheme on a 16x16 grid with Gaussian beams each of total width 5. (Top row) Radiation intensity after $t \in \{3, 4, 5\}$ light crossing times along the longest grid dimension, showing no non-physical collision effects. (Bottom row) Fractional error in radiation intensity relative to exact traces of beam paths through the grid.}
    \label{fig:crb_large_gaussian}
\end{figure*}

The results of crossing radiation beam simulations after $t \in \{3, 4, 5\}$ light crossing times along the longest grid dimension, as well as the error relative to an exact state, are shown in Figures \ref{fig:crb_small_uniform}, \ref{fig:crb_large_uniform}, and \ref{fig:crb_large_gaussian}. In the first two figures, each beam has radiation intensity exactly equal to 1 at each point on the grid concentrated only along the 45 degree angle; the beams are centered at $L_x/4$ and $3L_x/4$, and the widths of the beams are chosen differently for each simulation but equal for both the left and right beams within a simulation. In the third figure, we maintain mostly the same setup except we generate the beams using a Gaussian profile according to the function $I(x) = \exp\left[-(x-x_0)^2/W^2\right]$, where $x_0$ is the beam center (chosen to be $L_x/4$ and $3L_x/4$ for the two beams) and $W$ is the full-width of the Gaussian distribution (chosen differently for each simulation). The fractional error for each plot is simulated by comparing the simulated to the expected radiation intensity point-by-point.

By simulating multiple beam crossing occurrences at the same points, we have shown that our code does not suffer the same shortcomings faced by moment methods in such scenarios. The radiation from the two beams do not collide or deform, as seen in Ref.~\cite{Asahina2020}. There is also no dispersion in the beamwidth for a unidirectional beam as seen in Refs. \cite{Davis2012} and \cite{Jiang2014}, which also means the radiation intensity necessarily remains constant throughout the beams rather than diffusing. The error plots show perfect error outside the beams and noisy error across the beams; we do note that the error decreases at the crossings; this is likely due to the central limit theorem as there is more radiation being sampled from each of those points, and so the variance in that region decreases substantially.

\section{\label{sec:conclusion} Discussion and conclusion}

Lattice Boltzmann methods are very attractive candidates for quantum radiative transfer simulations due to their strong structure and the ease of formulating the algorithm steps as quantum operations. However, our tests of traditional LBMs in non-scattering media demonstrate that they are not suitable for astrophysics simulations due to inevitable anisotropy. Our novel angular redistribution method, compatible with both classical and quantum LBMs, is able to fix this anisotropy so that it no longer constrains simulation accuracy. Furthermore, the quantum formulation requires a modest number of qubits, opening the door to a new scale of simulations which can model larger and more detailed realizations of the universe.

There are variety of ways in which the algorithm can be improved. For example, there is significant room for improvement in how the quantum circuits are constructed \textit{prior} to the simulation itself. Reducing the gate count for each step's quantum circuit would significantly speed up the algorithm runtime.
Moreover, it would reduce the intermediate classical cost of preparing the quantum circuits before they are sent to hardware. Once a target quantum architecture has been identified, identifying hardware-tuned gate decompositions may significantly reduce the gate cost.
Currently, our hardware-agnostic transpilation shows that approximately most of the gate cost is due to the AR circuit and propagation circuits.
The isometry computation necessary for converting the matrix to a controlled operator is one of the most expensive parts since it involves complex multi-controlled gates, but work has been done in the literature to optimize this computation for sparse matrices \cite{Iten2016,Malvetti2021}. Since we anyways expect there to only be $2^{n_l}$ non-zero entries in any propagation matrix of the $(2^{n_l})^2$ total entries, our operator can be considered highly-sparse and thus may admit even simpler representations.

The primary strategies we foresee in optimizing the time complexity of algorithm are (i) decreasing the complexity of each individual step and (ii) unifying multiple steps into a single step. The former is already handled in part by utilizing circuit optimization in Qiskit; the latter would require more formulaic modifications to the algorithm. For example, the authors of Ref.~\cite{Lee2025} perform photon interaction computations during their quantum walk propagation step.

One extension of the angular redistribution method itself is to redistribute the radiation along multiple adjacent directions in order to further reduce anisotropy. However, this will necessarily come at the cost of having worse shadows since radiation is being artificially spread further, and thus must be implemented with caution. Such a change may be most appropriate, for example, with higher-angular resolution lattice Boltzmann schemes, for which the angular sectors associated with each discrete direction are already smaller. In this case, the additional shadowing caused by higher-order angular redistribution would not be as significant because each angular redistribution step would disperse the radiation by a shallower angle.

Cosmological simulations also typically involve other processes such as hydrodynamics and chemistry, so it is also worth extending the lattice methodology used here for radiative transfer to these other problems. By better utilizing quantum processors, one can reduce interprocess communication with classical processers, further accelerating simulations.
 
The runtime complexity of a LBM simulation depends primarily on grid size but notably not the number of sources nor the quantity density of the simulation, at the cost of simulating empty regions with the same resolution as high-density regions. This issue could be solved, for example, by adapting the resolution of the grid to the density of important lattice quantities (in our case, intensity), a method known as adaptive mesh refinement \cite{Wise2011}, and found a way to utilize this with minimal cost. Regardless, because quantum LBMs have very low cost associated with larger grids, the overhead from this is affordable.

It is also necessary to run this algorithm on real quantum hardware to gauge its performance on near-term noisy devices. On these devices, it may be necessary to apply quantum error mitigation or correction methods in order to obtain reliable simulation results. Such testing is essential in revealing what further algorithmic improvements need be made.

The current study opens a new path towards useful quantum computation, where memory limitations are a primary concern. Once integration with other physical processes has been handled, full simulations can be run to study cosmology at a much larger scale.

\vspace{0.5cm}

\begin{acknowledgments}

We acknowledge funding support from NSF grant AST-2510197 and NASA grant 80NSSC21K1053. R.D. acknowledges the use of AI tools for literature review and debugging code; the authors formulated the algorithm and the angular redistribution step. This research was supported in part through research cyberinfrastructure resources and services provided by the Partnership for an Advanced Computing Environment (PACE) at the Georgia Institute of Technology, Atlanta, Georgia, USA.

\end{acknowledgments}

\textbf{Data availablil}


\bibliography{main}

\end{document}